\documentclass[manuscript]{aastex62}

\usepackage{amsmath} 
\usepackage{amssymb}
\usepackage{bm}
\usepackage{graphicx}
\usepackage{afterpage}
\usepackage{tikz}
\usepackage{eso-pic}
\usepackage{capt-of}
\usepackage{mathrsfs} 
\usepackage{hyperref}
\usepackage{upgreek}
\usepackage{verbatim}
\usepackage{color}

\newcommand{\zstr}{z_\mathrm{tr}}
\newcommand{\zsobs}{z_\mathrm{obs}}
\newcommand{\pstr}{\vec{p}_\mathrm{\;tr}}

\newcommand{\psobs}{\vec{p}_\mathrm{\;obs}}

\newcommand{\cfp}{\mathrm{C}_\mathrm{fp}}
\newcommand{\ttr}{\theta_\mathrm{tr}}

\newcommand{\tobs}{\theta_\mathrm{obs}}

\newcommand{\code}[1]{\texttt{#1}}

\newcolumntype{M}[1]{>{\centering\arraybackslash}m{#1}}

\makeatletter
\def\parsecomma#1,#2\endparsecomma{\def\page@x{#1}\def\page@y{#2}}
\tikzdeclarecoordinatesystem{page}{
    \parsecomma#1\endparsecomma
    \pgfpointanchor{current page}{north east}
    \pgf@xc=\pgf@x%
    \pgf@yc=\pgf@y%
    \pgfpointanchor{current page}{south west}
    \pgf@xb=\pgf@x%
    \pgf@yb=\pgf@y%
    \pgfmathparse{(\pgf@xc-\pgf@xb)/2.*\page@x+(\pgf@xc+\pgf@xb)/2.}
    \expandafter\pgf@x\expandafter=\pgfmathresult pt
    \pgfmathparse{(\pgf@yc-\pgf@yb)/2.*\page@y+(\pgf@yc+\pgf@yb)/2.}
    \expandafter\pgf@y\expandafter=\pgfmathresult pt
}
\makeatother

\begin{document} 

\title{A Framework for Measuring Weak-Lensing Magnification Using the Fundamental Plane} 

\date{\today}

\author[0000-0001-6790-2939]{Jenna K.C. Freudenburg}
\affiliation{Center for Cosmology and AstroParticle Physics (CCAPP), The Ohio State University, Columbus, Ohio 43210, USA}
\affiliation{Department of Astronomy, The Ohio State University, 140 West 18th Avenue, Columbus, Ohio 43210, USA}

\author[0000-0002-9378-3424]{Eric M. Huff}
\affiliation{Jet Propulsion Laboratory, California Institute of Technology, 4800 Oak Grove Dr., Pasadena, CA 91109, USA}

\author[0000-0002-2951-4932]{Christopher M. Hirata}
\affiliation{Center for Cosmology and AstroParticle Physics (CCAPP), The Ohio State University, Columbus, Ohio 43210, USA}
\affiliation{Department of Physics, The Ohio State University, 191 West Woodruff Avenue, Columbus, Ohio 43210, USA}
\affiliation{Department of Astronomy, The Ohio State University, 140 West 18th Avenue, Columbus, Ohio 43210, USA}

\begin{abstract} 
Galaxy-galaxy lensing is an essential tool for probing dark matter halos and constraining cosmological parameters. While galaxy-galaxy lensing measurements usually rely on shear, weak-lensing magnification contains additional constraining information. Using the fundamental plane (FP) of elliptical galaxies to anchor the size distribution of a background population is one method that has been proposed for performing a magnification measurement. We present a formalism for using the FP residuals of elliptical galaxies to jointly estimate the foreground mass and background redshift errors for a stacked lens scenario. The FP residuals include information about weak-lensing magnification $\kappa$, and therefore foreground mass, since to first order, nonzero $\kappa$ affects galaxy size but not other FP properties. We also present a modular, extensible code that implements the formalism using emulated galaxy catalogs of a photometric galaxy survey. We find that combining FP information with observed number counts of the source galaxies constrains mass and photo-z error parameters significantly better than an estimator that includes number counts only. In particular, the constraint on the mass is 17.0\% if FP residuals are included, as opposed to 27.7\% when only number counts are included. The effective size noise for a foreground lens of mass $M_H=10^{14}M_\odot$, with a conservative selection function in size and surface brightness applied to the source population, is $\sigma_{\kappa,\mathrm{eff}}=0.250$. We discuss the improvements to our FP model necessary to make this formalism a practical companion to shear analyses in weak lensing surveys.
\end{abstract}


\section{Introduction} \label{sec:intro}

One of the key questions that confronts cosmology today is whether the $\Lambda$CDM model provides a correct, self-consistent description of the universe we live in. Precision cosmology attempts to answer this question by using multiple complementary cosmological probes to pin down the value of fundamental cosmological parameters \citep{Weinberg:2013aa}. Weak gravitational lensing has proved vital piece of this puzzle, providing invaluable constraints on the amplitude of structure and hence on cosmic acceleration \citep{Heymans:2013aa, Huff:2014ab, Troxel:2017aa, Prat:2017aa, Hildebrandt:2017aa,  Kohlinger:2017aa, DES-Collaboration:2017aa, Hikage:2019aa, Hamana:2019aa}. Furthermore, weak lensing has found a rich variety of applications, from constraining neutrino masses \citep{Abazajian:2011aa,DES-Collaboration:2017aa}, to probing dark matter halos and their connection to galaxy properties \citep{Brainerd:1996aa, Hudson:1998aa, Hoekstra:2004aa, Mandelbaum:2006aa, Mandelbaum:2016aa}. 

The most widely-used weak lensing statistics involve shear, the component of lensing that induces correlations in the distortion of background galaxy ellipticities. These statistics include shear-shear correlations (cosmic shear) and galaxy-shear correlations (galaxy-galaxy lensing). Producing high-quality measurements of cosmic shear and galaxy-galaxy lensing, among other lensing statistics, is a major goal of large-scale photometric galaxy surveys, including as the Sloan Digital Sky Survey (SDSS) \citep{Alam:2015aa}, the Dark Energy Survey (DES) \citep{Abbott:2018aa}, the Kilo-Degree Survey \citep{de-Jong:2017aa}, and the more targeted Hyper Suprime-Cam survey \citep{Aihara:2018aa}. In order for shear measurements to provide useful cosmological constraints, the methodology used to translate galaxy images from photometric surveys into such shear measurements must have a rigorous framework for dealing with systematic errors. Recently, significant effort has gone into thoroughly characterizing the effect of both astrophysical systematics, such as intrinsic alignments \citep{Croft:2000aa, Heavens:2000aa, Catelan:2001aa, Crittenden:2001aa, Hirata:2004aa, Hirata:2007aa, Mandelbaum:2011aa, Joachimi:2011aa, Heymans:2013aa, Troxel:2015aa, Krause:2016aa} and nonlinear and baryonic effects on the power spectrum \citep{Levine:2006aa, Rudd:2008aa, Heitmann:2009aa, van-Daalen:2011aa, Semboloni:2011aa, Zentner:2013aa, Heitmann:2014aa, Eifler:2015aa, Mohammed:2018aa, Wibking:2018aa}; and observational systematics, such as shear calibration bias \citep{Erben:2001aa, Hirata:2003aa, Heymans:2006aa, Massey:2007aa, Mandelbaum:2012aa, Mandelbaum:2015aa, Huff:2017aa, Sheldon:2017aa} and photometric redshift estimation \citep{Huterer:2006aa, Newman:2008aa, Hildebrandt:2010aa, Sanchez:2014aa, Masters:2017aa, Tanaka:2018aa}. An even deeper understanding of shear systematics (those mentioned above and many others - see \citet{Mandelbaum:2017aa} for a comprehensive account) will be necessary to take full advantage of the next generation of galaxy surveys, which will include the Large Synoptic Telescope Survey \citep{LSST-Science-Collaboration:2009aa}, the Wide Field Infrared Survey Telescope \citep{Spergel:2015aa}, and Euclid \citep{Amendola:2018aa}. As these surveys push to greater depth and achieve higher signal-to-noise than current surveys, we will need new methods for constraining both well-understood and heretofore-ignored systematics.

Weak-lensing magnification -- that is, the isotropic size increase or decrease induced by a lensing potential on a background galaxy image -- provides a useful path towards better constraints on lensing systematics. Different systematic effects apply to magnification and shear measurements, which suggests that the two may constrain each other in mutually useful ways. For example, shear measurement is strongly sensitive to the ellipticity of the point spread function (PSF); in the presence of astigmatism, this varies with focus position even for small perturbations around the position of best focus \citep{Born:1999aa, Noecker:2010aa}, which has been a major issue in weak lensing PSF modeling \citep{Jarvis:2004aa}. Magnification relies instead on the size of the PSF, which is minimized (hence has zero derivative) at the position of best focus. Hence we expect magnification to be less sensitive to focus drift than shear, and in general it will have different sensitivities to PSF systematics.

Although magnification has historically been the neglected counterpart of shear in the development of weak lensing frameworks, we are not the first to notice their complementary potential (see, e.g., \citealt{van-Waerbeke:2010aa}). The magnification signal has been detected alongside shear, most notably in SDSS \citep{Scranton:2005aa} by exploiting quasar number counts, and more recently in the DES Science Verification data \citep{Garcia-Fernandez:2018aa} using galaxy number counts. However, in neither case is the detection sufficient to provide useful cosmological constraints in combination with shear. \citet{Menard:2010aa} use quasar counts from SDSS to derive a galaxy-mass correlation that can be usefully compared with the galaxy-mass correlation inferred from shear \citep{Sheldon:2004aa}. Targeting Lyman-break galaxies has also proved successful in detecting a magnification signal at high significance \citep{Morrison:2012aa}. Moving beyond number counts, \citet{Schmidt:2012aa} obtain a magnification measurement by exploiting the sizes and magnitudes of galaxies around a sample of X-ray-selected clusters. In spite of these notable examples, progress toward a comprehensive magnification framework that can complement shear measurements has been slow.

In recent years, a number of authors have suggested use of the Fundamental Plane (FP) as a promising new approach to magnification measurement \citep{Bertin:2006aa,Huff:2014aa}. This method exploits the well-attested fact that early-type galaxies adhere with remarkably low scatter to a set of scaling relations among size, surface brightness, and stellar velocity dispersion. The primary attraction of the FP method is that it does not rely on number counts, which suffer a partial cancellation between the increased flux of galaxies at $\kappa>0$ and the reduced number density due to the change in the area element (see equation 64 of \citealt{Weinberg:2013aa}). Instead, the FP offers a (noisy) estimate for magnification at the location of each individual galaxy. Given the depth and volume of current and future surveys, we expect that even with a selection of only early-type galaxies, the sample size of lensed images will allow a vastly improved signal-to-noise over detections made with number counts.

In this paper, we model a simultaneous measurement of the foreground mass and redshift error distribution that exploits information about magnification contained in the fundamental plane residuals of each galaxy. In section \ref{sec:estimator} we define a generalized maximum likelihood estimator that recovers the input parameters (such as lens mass and redshift error distribution) given the observables generated by a model such as the one we define in section \ref{sec:physicalmodel}. In constructing this model, which accounts for the impact of weak lensing and redshift errors on photometric galaxy properties analogous to the traditional FP properties, we have in mind a survey similar to DES. Section \ref{sec:testing} describes the emulation code we have written to implement and test our estimation procedure on mock survey data generated using our model. In section \ref{sec:results} we present and discuss the results of applying this code to a fiducial case, before considering next steps in section \ref{sec:conclusions}.

\vspace{0.5cm}
\section{Fundamental Plane Overview}\label{sec:fp}

The physical properties of galaxies are determined by a wide variety of factors, but in the simple case of an elliptical galaxy with minimal dynamics, we expect the size, surface brightness, and stellar velocity dispersion to be related via the virial theorem:
\begin{equation}
    \langle v^2\rangle \propto \frac{GM}{R} \propto R\times \frac{L}{R^2}\times \frac{M}{L}.
\end{equation}
Indeed, we observe a that color-selected elliptical galaxies exhibit a tight correlation in the parameter space defined by size, surface brightness, and stellar velocity dispersion \citep{Bernardi:2003aa}. The plane on which this correlation lies, along with the associated scaling relations, is known as the fundamental plane (FP). 

Under this definition, the fundamental plane is, strictly speaking, only obtainable through spectroscopic measurements. However, we can define an analogous relationship among photometric quantities:
\begin{itemize}
\item size $\log_{10} R$, as inferred via angular size and a photometric redshift $\zsobs$;
\item surface brightness $\mu$, defined as $\log_{10} (F/T)$, where $F$ is flux in arbitrary linear units (defined $F=10^{(30-\mathrm{magnitude})/2.5}$) and $T$ is the size of the image in square arcsec; and
\item a parameter corresponding to the concentration of the light profile, which we will denote $c$.
\end{itemize}
We note that of these three properties, only size is ideally affected by magnification in the weak lensing regime\footnote{In practice, concentration obtained from a model fit may depend on magnification if the model does not exactly fit the galaxy profile, or at finite S/N, because of changes in how errors in different parts of the profile are weighted. These are analogous to the model biases \citep[e.g.][]{Voigt:2010aa} and noise-rectification biases \citep{Kaiser:2000aa, Bernstein:2002aa, Kacprzak:2012aa} commonly discussed in shear measurement.}. In the presence of magnification, $\log_{10} R \rightarrow \log_{10}(1+\kappa)+\log_{10} R$. This effect can be derived directly from the standard weak lensing formalism of \citet{Bartelmann:2001aa} (e.g. in \citet{Weinberg:2013aa}).

Some intrinsic scatter exists in the fundamental plane. \citet{Bernardi:2003aa} find an rms dispersion of 0.05 in $\log_{10} R$ relative to the center of the spectroscopic FP for SDSS galaxies. This dispersion is in part attributable to the evolution of FP with redshift. Furthermore, it has been shown by \citet{Joachimi:2015aa} that the FP depends on galaxy environment: fitting the FP from satellite galaxies alone, or from field galaxies, yields a result systematically lower in size than fitting from the brightest galaxies in groups. 

Additional biases in the plane may arise from observational effects. One notable effect, highlighted by \citet{Martens:2018aa}, is the correlation of orientation with position on the FP. Galaxies aligned along the the line of sight are measured to have smaller size and higher surface brightness, while the opposite is true for galaxies aligned perpendicular to the line of sight (this may also introduce additional selection bias). We anticipate that a thorough understanding of this effect, as well as the evolution and environmental effects mentioned above, will be necessary for the FP method to reach its full potential, but we defer this question for future study.

The photometric FP is particularly susceptible to observational effects. For one thing, any convenient concentration parameter correlates imperfectly with the velocity dispersion and will be difficult to measure accurately even if well-chosen. More pernicious are the photometric redshift errors, which, in addition to impacting the inferred physical sizes of galaxies, bias the galaxy-size cross-correlation when galaxies clustered with foreground lenses are erroneously measured as background sources. Characterizing the photo-z error forms an important part of our analysis; however, we leave aside the question of photo-z clustering bias for the time being, since we do not fully realize a foreground lens population (see section \ref{sec:physicalmodel}).

With these considerations in mind, we write the fundamental plane relationship for a single elliptical galaxy as follows:
\begin{equation}
\kappa = \log_{10} R - \big[ a(\zstr)\times\mu + b(\zstr)\times c + q_z(\zstr) + q_M(\zstr)\log_{10} M_H \big].
\end{equation}
Here, $q_z$ parameterizes the evolution of the FP with redshift, and $q_M$ (which may itself vary with redshift) parameterizes evolution with environment, for which the mass of the host halo $\log_{10} M_H$ is a proxy. For the purposes of the code demonstration in this paper, we set $q_z$ and $q_M$ to 0, although we plan in the future to more fully model variations in the FP and have built the capability to do so into our emulator code.

We have already alluded to the fact that choosing a suitable concentration parameter requires some thought. One could, for example, use the S\'ersic index \citep{Graham:2002aa} or a ratio of radii enclosing different percentiles of the total flux (as in \citealt{Huff:2014aa}), but this requires data from a survey pipeline that can accurately discern higher-order moments of the light profile at finite $S/N$ ratio. As experience with shear analyses has taught us, this is no trivial requirement; {\em any} concentration parameter depends on at least the 4$^{\rm th}$ moment of the light distribution. Alternatively, we can use a projected version of the FP, and accept the additional scatter introduced by the tilt of the FP relative to the $(\mu,\log_{10} R)$ plane. This is the option we choose for our analysis.

\vspace{0.5cm}
\section{Estimator Formalism} \label{sec:estimator}
Here we derive a mathematical formalism for estimating the true values of quantities underlying a galaxy population, given a model for generating observables from these quantities. 
\subsection{Minimal working example}
Let's start with a basic scenario: assume we are ideal observers along a line of sight with a single lens in the foreground and an arbitrary number of source galaxies in the background. Because we are ideal observers, the observables to which we have access are (a) true redshifts and (b) galaxy properties that differ from the true properties due only to the effects of gravitational lensing. Furthermore, assume we have perfect knowledge of the fundamental plane, such that we have its true covariance matrix $\cfp$. Define a vector of galaxy parameters $\vec{p}=(\log_{10}R,\mu,c)$. Then, given the conditions above, we may stipulate that for any given galaxy, $\vec g_\mathrm{ideal}=\vec{p}_\mathrm{ideal}-\pstr$ corresponds exactly to the effects of magnification. We may write the log likelihood of $\vec g_\mathrm{ideal}$ as
\begin{equation}
\ln L = -\frac{1}{2}\left(\vec g_\mathrm{ideal} - \kappa\hat\delta\right)^T\,\cfp^{-1}\,\left(\vec g_\mathrm{ideal} - \kappa \hat\delta\right) + {\rm constant},
\end{equation}
where $\hat\delta = (1/\ln(10),0,0)$ since magnification affects neither size nor concentration. Then the value of $\kappa$ that maximizes the likelihood is given by
\begin{equation}
 \hat\kappa = \frac{\hat\delta^T\cfp^{-1}\vec{\emph g}_\mathrm{ideal}}{\hat\delta^T\cfp^{-1}\hat\delta}.
\end{equation}
(Note that this result holds for any $\hat\delta=(\lambda,0,0)$ where $\lambda$ is a constant, although $\lambda = 1/\ln 10$ is the most physical choice in this case.) In our idealized, systematics-free scenario, this expression is best understood as the optimal $\kappa$ estimator for an individual galaxy, which gives us direct access to the surface mass density along the line of sight. 

In a more realistic scenario, we don't have access to either the true or the systematics-free properties. In this case, we can write a similar expression 
\begin{equation}\label{eqn:residual}
 \Delta = \frac{\hat\delta^T\cfp^{-1}\vec{\emph g}}{\hat\delta^T\cfp^{-1}\hat\delta}.
\end{equation}
where $\vec g=\psobs-\overline\psobs$ for a single galaxy. $\Delta$ should be understood as the fundamental plane residual, i.e. the optimal estimator for the offset of that galaxy from the fundamental plane, whether due to magnification or other factors\footnote{If the user wishes to incorporate a model in which galaxy properties beyond size are impacted by magnification (e.g. due to model biases), this definition and the subsequent formalism may be naturally extended by modifying $\hat\delta$ accordingly.}. In this case, since we do not have prior knowledge of the ``true'' FP, $\cfp$ and $\overline\psobs$ must be fit from the full population (in the absence of significant selection effects, which address in section \ref{ssec:truncated_distributions}).

\subsection{Generalized estimator}

Suppose we have a model $M(\Theta):\mathbb{R}^n \rightarrow\mathbb{R}^m$ that a takes a vector $\Theta$ of $n$ model parameters and returns a data vector $d$ with $m$ elements. For our analysis, we choose parameters $\Theta$ that include lens mass, redshift distribution, cosmology, and others. The full set of parameters, along with their fiducial values, are shown in table \ref{tab:fiducial_values}. The estimator also requires a data vector $d$, which could theoretically contain one or more entries for each galaxy. However, we instead choose to define summary statistics in bins of redshift and angular position around the foreground lens (bins given in table \ref{tab:settings}). The first half of our data vector comprises the number density of galaxies in each bin, and the second half averages the residuals in each bin (mean $\Delta$) weighted by bin area. Section \ref{sec:physicalmodel} describes our observables model analytically; ultimately, we implement this model with our emulator code, which generates a model data vector for a set of input parameters $\Theta$, as described in section \ref{sec:testing}. The covariance matrix C for the data vector (whose calculation is also detailed in section \ref{sec:testing}) then has dimensions $m\times m$, where $m = 2\times(\mathrm{number\;of\;}z\mathrm{\;bins})\times(\mathrm{number\;of\;\theta\;bins})$. Note that in spite of the specific choices detailed here, what follows is a highly flexible formalism that can accommodate any configuration for $\Theta$, $M$, and $d$.

Given a set of input parameters, a model, a data vector, and a covariance matrix, we can maximize the likelihood of the data vector $d$ by solving the system of equations
\begin{equation}
\frac{\partial \ln L}{\partial\Theta_i} = -\frac{1}{2}\frac{\partial}{\partial\Theta_i}\left[\big(d_a-M_a(\Theta)\big)C^{-1}_{ab}\big(d_b-M_b(\Theta)\big)\right] = 0,
\end{equation}
where $a,b,\in\mathbb{R}^m$, $i,j\in\mathbb{R}^n$, and the Einstein summation convention is applied. Assuming that the covariance is relatively robust to variations in the parameters, we may simplify:
\begin{equation}\label{eq:maxlike}
-\frac{\partial M_a}{\partial\Theta_i}\;C^{-1}_{ab}\left(d_b-M_b(\Theta)\right) = 0.
\end{equation}
With some manipulation, this equation can be linearized. We can Taylor expand $M_a(\Theta_0)$ around a fiducial parameter vector $\Theta_0$:
\begin{equation}
M_a(\Theta) \approx M_a(\Theta_0)+(\Theta_i-\Theta_{i,0})\left.\frac{\partial M_a}{\partial \Theta_i}\right\vert_{\Theta_0}.
\end{equation}
Each partial-derivative component may also be linearly approximated, using the absolute change across the fiducial parameters $B_a$ as defined below:
\begin{equation}\label{eqn:linear_response}
\left.\frac{\partial M_a}{\partial \Theta_i}\right\vert_{\Theta_0} \approx \frac{1}{2\delta\Theta_i}\left(M_a(\Theta_0+J_i\delta\Theta) - M_a(\Theta_0-J_i\delta\Theta)\right) \equiv \frac{B_a}{2\delta\Theta_i}.
\end{equation}
Here $J_i$ is the single-entry $n\times n$ matrix for which the $i$th diagonal entry is 1. We also make the definition $A_a \equiv d_a-M_a(\Theta_0)$. Then we can rewrite equation \ref{eq:maxlike} as follows (notated with explicit summation for clarity, although we continue to apply the Einstein convention in general):
\begin{equation}
\frac{B_a}{2\delta\Theta_i}C^{-1}_{ab}\Bigg(A_b-\sum\limits_{j=1}^n(\Theta_j-\Theta_{j,0})\; \frac{B_b}{2\delta\Theta_j} \Bigg) = 0.
\end{equation}
This system has the solution
\begin{equation}\label{eqn:estimator}
\hat\Theta_j - \Theta_{j,0} = \Bigg[\frac{B_a}{2\delta\Theta_i} C^{-1}_{ab} \bigg(\frac{B_a}{2\delta\Theta_j}\bigg)^T \,\Bigg]^{-1}_{ij} \frac{B_a}{2\delta\Theta_i}C^{-1}_{ab} A_b,
\end{equation}
which gives the optimal estimator for the model parameters.

\subsection{Selection effects and truncated distributions}\label{ssec:truncated_distributions}
When selection in $(\mu,\log_{10} R)$ space is applied to a galaxy population, the normalization of the log-likelihood is not a constant, so we must add an additional term to the derivative. 

Assume a selection function that divides the fundamental plane galaxies into populations A (selected galaxies) and B (non-selected galaxies). Then we have $P(\mathrm{gal.\;in\;A})+P(\mathrm{gal.\;in\;B})\equiv P_A+P_B=1$, so $P_A$ is the normalization constant for the likelihood, i.e.
\begin{equation}\label{eqn:likelihood_selection}
    \ln L = -\frac{1}{2}(\vec{g}-\Delta\hat\delta)^T\cfp^{-1}(\vec{g}-\Delta\hat\delta) - \ln P_A
\end{equation}
where $\Delta$ is the vertical-axis fundamental plane residual. Therefore
\begin{equation}
    \frac{\partial \ln L}{\partial \Delta} = -\frac{1}{2}\hat\delta\;\cfp^{-1}(\vec{g}-\Delta\hat\delta) - \frac{\partial P_A}{\partial \Delta}\frac{1}{P_A}.
\end{equation}
Solving for the maximum likelihood, then, we have
\begin{equation}\label{eqn:dpdDelta_term}
    \Delta = \frac{2P_A^{-1}\times\partial P_A/\partial \Delta + \hat\delta^T\cfp^{-1}\vec{g}}{\hat\delta^T\cfp^{-1}\hat\delta}.
\end{equation}
The additional term in the numerator can be approximated
\begin{equation}
    \frac{\partial P_A}{\partial \Delta}\frac{1}{P_A} \approx \frac{1}{N_A|_\Delta}\,\frac{N_A|_{\Delta+\delta\Delta}-N_A|_{\Delta-\delta\Delta}}{2\delta\Delta}
\end{equation}
where $N_A$ is the number of galaxies in the sample that pass selection and $\delta\Delta$ is a small perturbation. 
In cases where selection effects are significant, we cannot rely on fitting the FP directly from our observed sample, so we must have another means of determining $\overline\psobs$ and $\cfp$. For example, in DES, we expect to obtain the FP fit from the deep field observations, which will yield a much more complete galaxy sample than the main survey.

\section{Physical Model} \label{sec:physicalmodel}

We posit a stacked lens configuration, in which a single foreground lens of mass $M_H$ is superimposed on a background population of source galaxies with number density proportional to the expected number of foreground lenses of that mass. In this section we detail our modeling choices for such a scenario. In order to implement the estimator described in section \ref{sec:estimator}, we model the observables $\psobs$ (observed galaxy properties), $\zsobs$ (observed redshift), and $\tobs$ (observed angular separation) for each source galaxy. Fiducial values for the parameters used in this model are shown in table \ref{tab:fiducial_values}. 

\begin{table}[p!]
    \centering
    \begin{tabular}{|c|c|c|c|}
    \hline
Parameter & Key & Description & Fiducial Value \\
\hline
\hline
$\{\Omega_m\}$,...         &                   & cosmology   & Planck 2013 values \\   
\hline
$z_L$                 & \code{lens\_z}    & redshift of the lens &  0.1 \\
$M_h$                 & \code{lens\_mass} & mass of the lens halo &  $10^{14} M_\odot$  \\
\hline
$\alpha$              & \code{alpha}      & see equation \ref{eqn:nz_intrinsic} &  1.3 \\
$\beta$               & \code{beta}       & see equation \ref{eqn:nz_intrinsic} &  1.0 \\
$z_0$                 & \code{z0}         & see equation \ref{eqn:nz_intrinsic} &  0.25 \\
$z_\mathrm{min}$      & \code{zmin}       & minimum of redshift range           &  0.0  \\
$z_\mathrm{max}$      & \code{zmax}       & maximum of redshift range           &  2.0  \\
\hline
$w$                   & \code{w\_coeff}    & see equation \ref{eqn:theta_distribution} & 3.0 \\
$\theta_0$            & \code{t0}         & see equation \ref{eqn:theta_distribution} & 36.0 arcsec\\
$m$                   & \code{m}          & see equation \ref{eqn:theta_distribution} & 0.7  \\
$\theta_\mathrm{min}$ & \code{tmin}       & minimum angular separation &  0.01 arcsec \\
$\theta_\mathrm{max}$ & \code{tmax}       & maximum angular separation &  300 arcsec \\
\hline
$\overline{\pstr}$ & \code{mean}  & mean of the galaxy properties &  (0.815 kpc, 4.02 flux units/arcsec$^2$) \\
$\cfp$                & \code{cov}        & covariance of the galaxy properties & $\begin{pmatrix} 0.0637 & -0.0673 \\ -0.0673 & 0.111 \end{pmatrix}$ \\
\hline
$\sigma_z$            & \code{pzerr\_std}       & amplitude of the redshift error &  0.02 \\
$\mu_{z,1}$           & \code{pzerr\_mean\_1}   & mean of the redshift error in bin 1 &  -0.001 \\
$\mu_{z,2}$           & \code{pzerr\_mean\_2}   & mean of the redshift error in bin 2 &  -0.019 \\
$\mu_{z,3}$           & \code{pzerr\_mean\_3}   & mean of the redshift error in bin 3 &  0.009 \\
$\mu_{z,4}$           & \code{pzerr\_mean\_4}   & mean of the redshift error in bin 4 &  -0.018 \\
\hline
$a_0$ & \code{selection\_intercept}	& selection boundary intercept & 33.0 \\
$a_1$   				  & \code{selection\_slope}  	& selection boundary slope & -8.0 \\
\hline
    \end{tabular}
    \caption{Parameters and fiducial values used in this analysis.}
    \label{tab:fiducial_values}
\end{table}

\subsection{Galaxy Distribution} \label{ssec:galaxy_distribution}
\begin{figure}[h]
    \centering
    \includegraphics[scale=0.8]{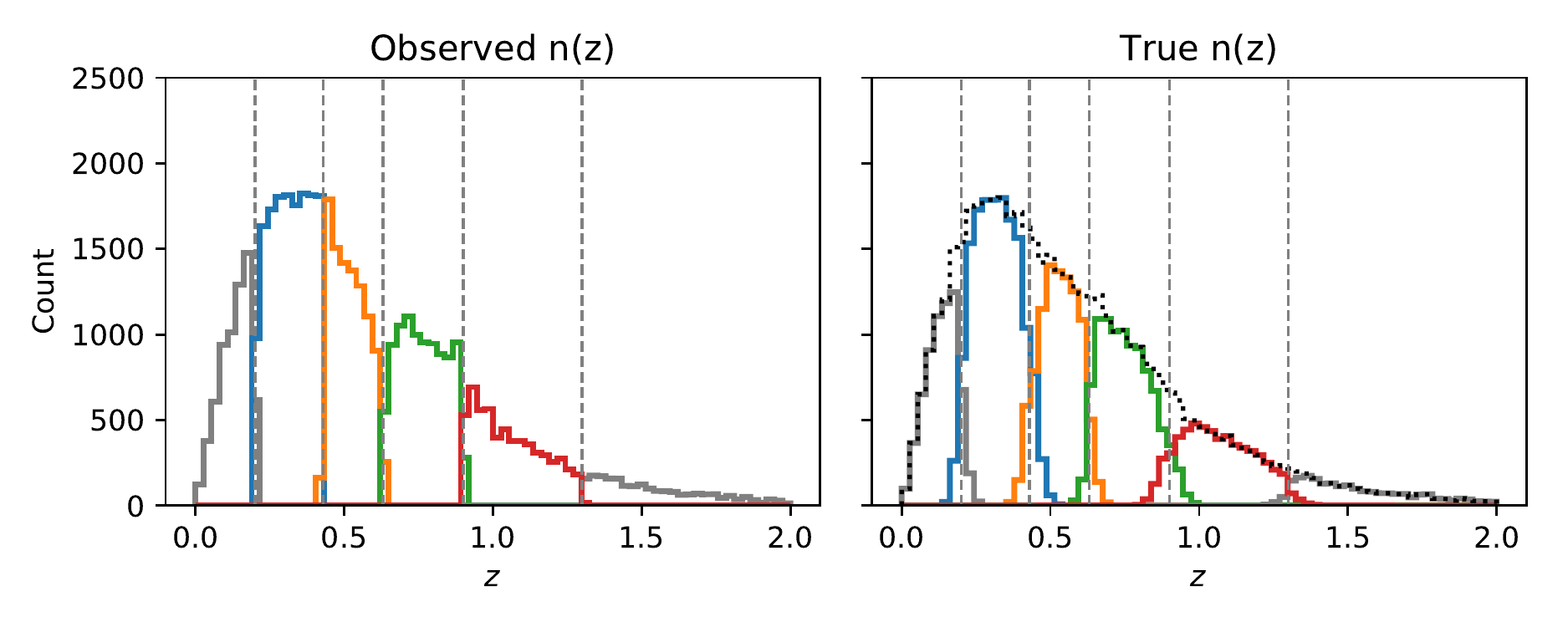}
    \caption{The true and observed redshift distributions generated by our redshift model from fiducial parameter values. Colors indicate redshift bins used in the analysis, while grey indicates parts of the distribution that lie outside the redshift bins of the main sample. The dotted black line is an eye guide for the full distribution. Bin boundaries are listed in table \ref{tab:settings}.}
    \label{fig:n_of_z}
\end{figure}

We parameterize the source galaxy distribution with the generalized gamma function
\begin{equation}\label{eqn:nz_intrinsic}
    n(z) = z^\alpha \exp\left[{-\left(\frac{z}{z_0}\right)^\beta}\right],
\end{equation}
where $z_0$, $\alpha$, and $\beta$ are free parameters and $z$ is the true redshift. Our fiducial values for these parameters are calibrated to produce a distribution resembling $n(z)$ for the simulated lensing sample in \citet{Hoyle:2018aa} (see their figure 5). The distribution of true galaxy redshifts produced by this parameterization is shown in the right panel of figure \ref{fig:n_of_z}.

\begin{figure}[h]
    \centering
    \includegraphics[scale=0.8]{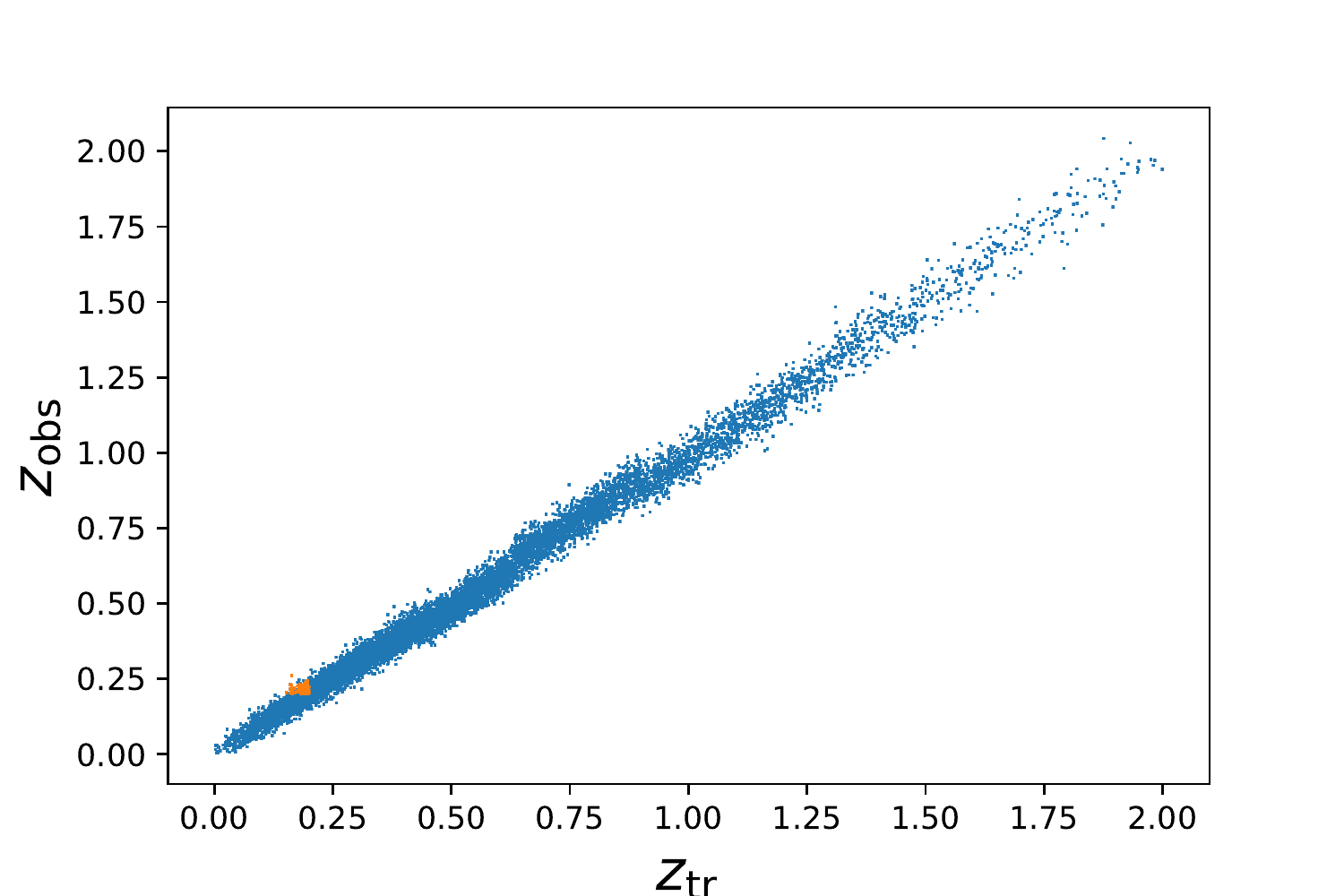}
    \caption{Relationship between the true and observed redshifts. The area highlighted in orange corresponds to galaxies that fall below the lowest redshift bin, and therefore cluster with the lens, but are scattered into the sample.}
    \label{fig:z_vs_z}
\end{figure}

Angular positions $\theta$ with respect to the lens are assigned randomly with a uniform distribution over an annular area defined by support $(\theta_\mathrm{min}$,$\theta_\mathrm{max})$. If source galaxies are in the redshift bin of the lens (i.e. falling below the range of the redshift bins shown in table \ref{tab:settings}), they are assumed to cluster with the lens according to the distribution
\begin{equation}\label{eqn:theta_distribution}
    f(\theta) = \theta\left[1+w\left(\frac{\theta}{\theta_0}\right)^{-m}\right],
\end{equation}
where $\theta\,d\theta$ is proportional to the area element in the flat-sky approximation and $w$ and $m$ parameterize the angular correlation function. For all bins listed above, and any source galaxies falling above these bins, $w=0$ and a uniform distribution in area results. For the lens bin, we choose $m=0.7$, in accordance with \citet{Connolly:2002aa}. As figure \ref{fig:z_vs_z} demonstrates, only a small number of clustered galaxies are present in the source population. 

In our particular fiducial context the modeling choice described above is somewhat inconsistent, since it allows galaxies just in front of the the lowest redshift bin ($z_S<\approx 0.2$) to be simultaneously clustered with and magnified by the lens at $z_L=0.1$, even though galaxies so far separated in redshift are not physically associated. However, our choice mimics the effect of a more realistic physical configuration, in which a fully realized foreground population has more distant galaxies, which may be clustered amongst themselves, that are lensed by nearer galaxies in the same foreground population and artificially scattered into the source population due to photo-z error. Capturing this effect and controlling the bias it introduces is an essential task for the FP method \citep{Huff:2014aa}. Full implementation of a foreground model will be survey-specific, however, and we leave it for future work.

\subsection{Lens \& Lensing}\label{ssec:lensing}

We assume a spherically symmetric lens with mass $M_H$, redshift $z_L$, and an NFW mass profile. The surface mass density $\Sigma$ is then given by the following (which matches equation 11 in \citet{Wright:2000aa}):
\begin{equation}
\Sigma_{\rm nfw}(x) = \left\{ \begin{array}{ll}
\frac{2r_{s}\delta_{c}\rho_{c}}{\left(x^{2}-1\right)}
\left[1-\frac{2}{\sqrt{1-x^{2}}}{\rm arctanh}\sqrt{\frac{1-x}{1+x}}\hspace{0.15cm}
 \right] 
& \mbox{$\left(x < 1\right)$} \\ 
 & \\
\frac{2r_{s}\delta_{c}\rho_{c}}{3} & \mbox{$\left(x = 1\right)$} \\ 
 & \\
\frac{2r_{s}\delta_{c}\rho_{c}}{\left(x^{2}-1\right)}
\left[1-\frac{2}{\sqrt{x^{2}-1}}\arctan\sqrt{\frac{x-1}{1+x}}\hspace{0.15cm}
 \right] 
& \mbox{$\left(x > 1\right)$}
\end{array}
\right.\,.
\end{equation}
Here $x$ is a dimensionless scale, $r_s x/\theta$ is the angular diameter distance to the lens, and $\delta_c,r_s$ and $\rho_c$ are mass- and halo concentration-dependent quantities (see the formalism of \citealt{Wright:2000aa} for definitions).

\begin{figure}[h]
    \centering
    \includegraphics[scale=0.8]{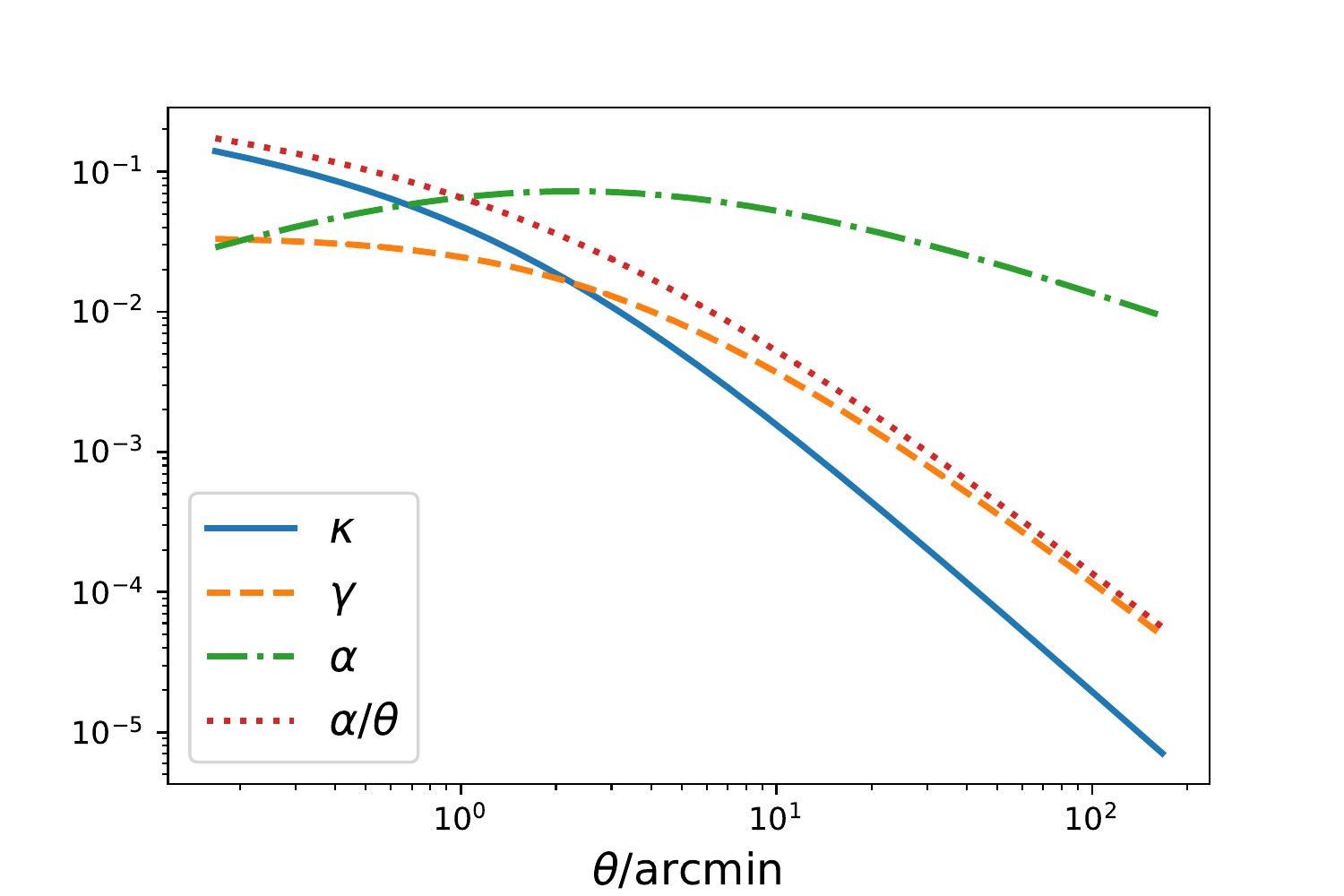}
    \caption{Profiles of lensing quantities from an idealized NFW halo with the fiducial values for halo mass and redshift.}
    \label{fig:kappagamma}
\end{figure}

The magnification for a source galaxy at redshift $z_S$ is given by 
\begin{equation}
\kappa(\theta) =
\begin{cases}
    \Sigma(\theta)/\Sigma_c, \hspace{10pt} z_S>z_L \\
     0, \hspace{45pt} z_S<=z_L\\
\end{cases}\,,
\end{equation}
where the critical density $\Sigma_c$ is
\begin{equation}
\Sigma_c = \frac{C^2}{4\pi G}\frac{D_1(z_S)}{D_2(z_L)D_{12}(z_L,z_S)}  .
\end{equation}
Here $C$ is the speed of light and $D_1,D_2$, and $D_{12}$ are the angular diameter distances at the source position, at the lens position, and between the source and lens positions. In calculating angular diameter distances, we adopt the central values of cosmological parameters $\{\Omega_m,...\}$ given in \citet{Planck-Collaboration:2014aa}. Figure \ref{fig:kappagamma} shows the $\kappa$ profile for our fiducial lens and includes the profiles for lensing displacement and shear for comparison. Shear is given by 
\begin{equation}
    \gamma(\theta) = 
    \begin{cases}
        \big(\overline\Sigma(\theta)-\Sigma(\theta)\big)/\Sigma_c, \hspace{10pt} z_S>z_L \\
        0,  \hspace{95pt} z_S<=z_L
    \end{cases}\,,
\end{equation}
where $\overline{\Sigma}(\theta)$ is the mean surface mass density interior to $\theta$. Per \citet{Golse:2002aa}, the lensing displacement $\alpha_\theta$ between the background galaxy position in the source plane and its position in the image plane is given by
\begin{equation}
    \alpha_\theta(\theta) = 
    \begin{cases}
    	\theta\times\overline{\Sigma}(\theta)/\Sigma_c, \hspace{10pt} z_S>z_L \\
		0, \hspace{67pt} z_S<=z_L
	\end{cases}\,.
\end{equation}
We observe the deflected position rather than the ``true" position of each galaxy; in section \ref{ssec:observables}, we discuss the impact of this effect on our observables. (Note that displacement $\alpha_\theta$ is different from redshift parameter $\alpha$.)

\subsection{Physical Properties}
\begin{figure}[h]
    \centering
    \includegraphics[scale=0.8]{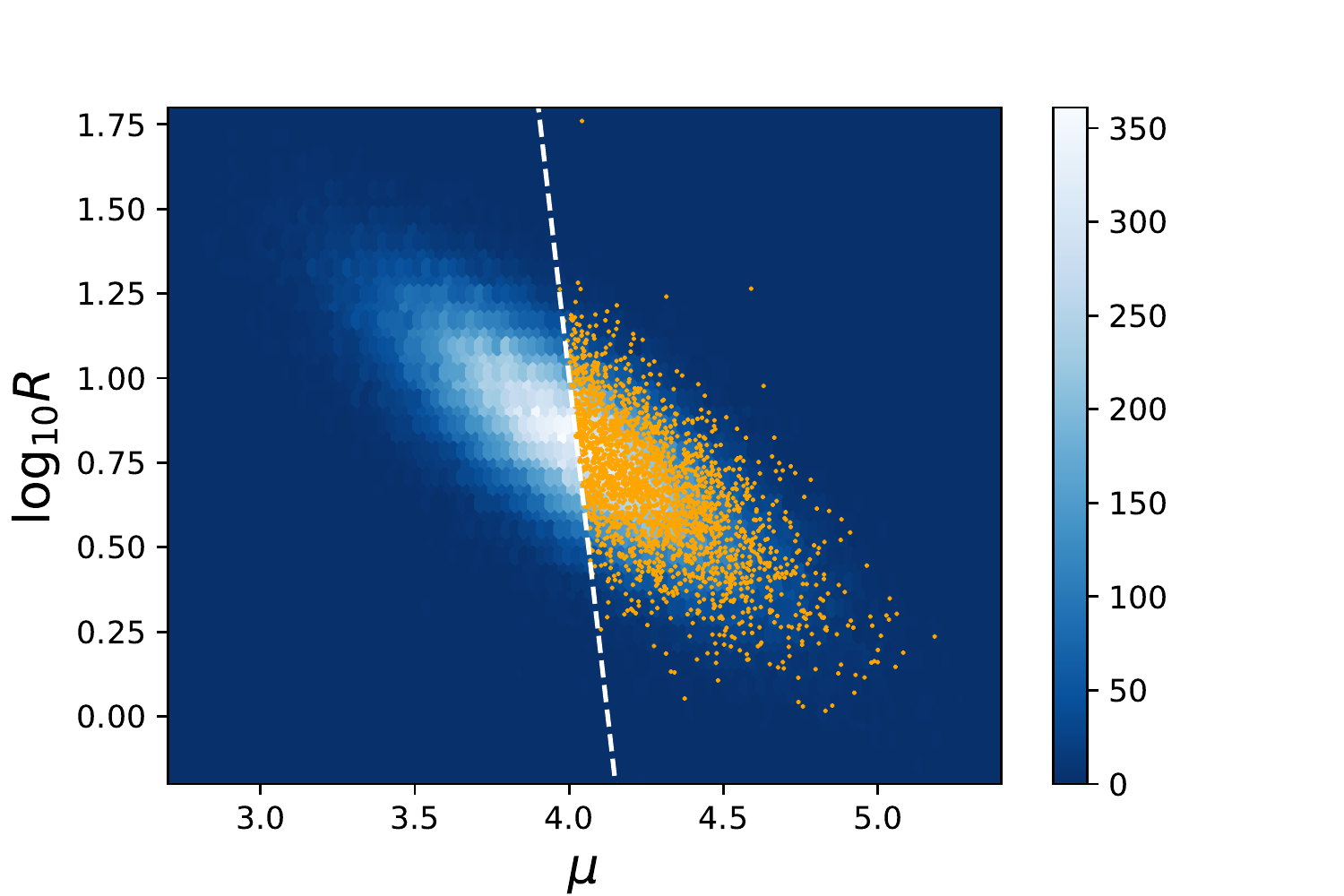}
    \caption{The fundamental plane of observed galaxy properties for all galaxies (blue and white density map) and for galaxies passing the selection cut (orange points, density reduced by factor of 20 for visual clarity). The dashed white line represents the selection boundary.}
    \label{fig:selection}
\end{figure}

As considered in section \ref{sec:fp}, we treat the plane in projection, using only $\log_{10} R$ and $\mu$ as our galaxy properties. We define a vector containing the true galaxy properties $\pstr=(\log_{10} R_\mathrm{tr},\mu_{tr})$ which is drawn for each source galaxy from the multivariate normal distribution 
\begin{equation}
P(\pstr)=\mathcal{N}(\overline\pstr,\cfp).
\end{equation}
For the purposes of this paper, we treat $\cfp$ and $\overline\pstr$ as free parameters instead of modeling them explicitly based on assumptions about the galaxy population. The model FP generated by our fiducial choices for $\cfp$ and $\pstr$ is shown in figure \ref{fig:selection}.

\subsection{Photometric Redshift Errors} \label{ssec:photoz_errors}
We follow the parameterization for redshift errors applied in \citet{Hoyle:2018aa}. For each galaxy, a redshift error $z_\mathrm{err}$ is drawn from the distribution
\begin{equation}
    p(z_\mathrm{err})=\mathcal{N}(\mu_{z,i},\sigma_z(1+z_S)),
\end{equation}
where $\mu_{z,i}$ is the mean of the redshift error in each bin and $\sigma_z$ is assumed to be constant across the whole population. Note that whereas \citet{Hoyle:2018aa} assign galaxies to redshift bins based on the expectation value of their posterior, we here assign galaxies to a bin based on their true values, since we are forward-modeling the redshift errors. The true and observed redshift distributions obtained from this prescription are shown in figure \ref{fig:n_of_z}. Figure \ref{fig:z_vs_z} shows the relationship between the true and observed redshifts over the an entire population. We note that while some galaxies with $\zstr \ll 1$ will have $\zsobs<0$ due to redshift error, these galaxies do not impact our analysis, since the source sample is selected to have $\zsobs>0.2$ through the data vector construction.

\subsection{Observables}\label{ssec:observables}
Once intrinsic quantities have been drawn from the appropriate distributions and lensing effects have been determined, we apply magnification, lensing displacement, and redshift error effects to produce the observed quantities $\zsobs$, $\tobs$, and $\psobs$. For redshift and angular position, we simply add:
\begin{equation}
    \zsobs = \zstr + z_\mathrm{err} ~~~{\rm and}~~~
    \tobs = \ttr + \alpha_\theta(\ttr) .
\end{equation}

The galaxy properties are impacted differently by redshift errors and magnification: size is subject to both effects, while surface brightness is conserved under magnification, changing only due to redshift errors because of the cosmological $(1+z)^4$ dimming effect. We make the simplifying assumption that all surface brightness quantities are bolometric, and therefore the effect of K-corrections may be safely ignored. We defer a more complete picture, including K-corrections and color-dependent effects more generally, to future work that incorporates a model for evolution of the FP. We assume that the $\mu$ column in our catalog has been corrected for surface brightness dimming. The uncorrected quantity may be written $\mu_\mathrm{tr}-4\log_{10}(1+\zstr)$. Then applying the correction inferred from the observed redshift, we have
\begin{equation}
    \mu_\mathrm{obs} = \mu_\mathrm{tr}-4\log_{10}\left(\frac{1+\zstr}{1+\zsobs}\right).
\end{equation}
The apparent size, $\log_{10} \left(R(1+\kappa)\right)$ is impacted via the angular diameter distance, such that \begin{equation}
    \log_{10} R_\mathrm{obs} = \log_{10} \left((1+\kappa)R_\mathrm{tr}\right) +  \log_{10}\left(\frac{D_a(\zsobs)}{D_a(\zstr)}\right).
\end{equation}

We may now write the full probability distribution for each of the observables given a parameter vector $\Theta$. For the redshift:
\begin{equation}
 \begin{split}
    P(\zsobs|\Theta) &= P(\zsobs|\zstr,z_\mathrm{err})P(z_\mathrm{err}|\mu_{z,i},\zstr,\sigma_z)P(\mu_{z,i}|\zstr)P(\zstr|\alpha,\beta,z_0,z_\mathrm{min},z_\mathrm{max}) \\
    &=\delta(\zsobs'-(\zstr+z_\mathrm{err}))\times\mathcal{N}(\mu_{z,i}(\zstr),(1+\zstr)\sigma_z)\times\frac{\int_{\mathrm{bin\;}i}n(z)dz}{\int_{z_\mathrm{min}}^{z_\mathrm{zmax}}n(z)dz}\times Bn(\zstr)
 \end{split}
\end{equation}
where $B$ is a normalization constant for the intrinsic redshift distribution and $\delta$ is the Dirac delta function. (Primed quantities within the $\delta$ functions are dummy variables). Similarly, for angular position:
\begin{equation}
 \begin{split}
    P(\tobs|\Theta) &=
    P(\tobs|\alpha_\theta,\ttr)  P(\alpha_\theta|\ttr,\Sigma,\Sigma_c) P(\ttr|w,\theta_0,m) \\%P(w|\zstr)\\
    &\hspace{15pt} \times P(\zstr|\alpha,\beta,z_0,z_\mathrm{min},z_\mathrm{max}) P(\Sigma|\zstr,\ttr,M_H) P(\Sigma_c|\zstr,z_L,\{\Omega_m,\mathrm{...}\}) \\
    &= \delta\left(\tobs'-(\alpha_\theta+\ttr)\right)\times\delta\left(\alpha_\theta'-\alpha_\theta(\ttr,\Sigma,\Sigma_c)\right)\times\delta(\Sigma'-\Sigma(\ttr,z_L,zs)) \\
    &\hspace{15pt} \times\delta(\Sigma_c'-\Sigma_c(\zstr,z_L,\{\Omega_m,\mathrm{...}\}))\times B n(\zstr)\times f(\ttr) %\\
%    &\hspace{15pt}\times 
%    \begin{cases}
%        w_\mathrm{fiducial} \hspace{15pt}\big(\zstr<\min(z \mathrm{bin}_1)\big) \\
%        0 \hspace{25pt}\big(\zstr>=\min(z \mathrm{bin}_1)\big)
%    \end{cases}.
 \end{split}
\end{equation}
Finally, the observed distribution of galaxy quantities may be written 
\begin{equation}
 \begin{split}
    P(\psobs|\Theta) &=
    P(\psobs|\zstr,\zsobs,\pstr,\kappa) P(\kappa|\Sigma,\Sigma_c) P(\ttr|w,\theta_0,m) \\%P(w|\zstr)\\
    &\hspace{15pt} \times P(\pstr|\cfp,\overline\pstr)P(\zsobs|\Theta) P(\Sigma|\zstr,\ttr,M_H) P(\Sigma_c|\zstr,z_L,\{\Omega_m,\mathrm{...}\}) \\
    &= \delta(\vec{p}\,'-\psobs(\pstr,\zstr,\zsobs))\times\delta(\kappa'-\Sigma/\Sigma_c)\times\delta(\Sigma'-\Sigma(\ttr,z_L,zs))\\
    &\hspace{15pt} \times\delta(\Sigma_c'-\Sigma_c(\zstr,z_L,\{\Omega_m,\mathrm{...}\}))\times\mathcal{N}(\overline\pstr,\cfp)\times f(\ttr)\times P(\zsobs|\Theta) %\\
    %&\hspace{15pt}\times 
    %\begin{cases}
    %    w_\mathrm{fiducial} \hspace{15pt}\big(\zstr<\min(z \mathrm{bin}_1)\big) \\
    %    0 \hspace{25pt}\big(\zstr>=\min(z \mathrm{bin}_1)\big)
    %\end{cases} .
 \end{split}
\end{equation}
Note that we understand $\psobs$ to indicate the galaxy properties as they \emph{would be} observed, whether or not an individual galaxy passes the selection cut.

\subsection{Selection}

We implement a hard selection boundary in $\log_{10} R$ and $\mu$, such that only galaxies with $\log_{10} R_\mathrm{obs}>a_0 + a_1 \mu_\mathrm{obs}$ are included in our source sample. While the mathematical framework in section \ref{ssec:truncated_distributions} accommodates any more generic selection function, this linear model (with our fiducial choices for $a_0$ and $a_1$) captures the cutoff in (mainly) surface brightness that we expect in a flux-limited survey. We choose for demonstration purposes a case where the cutoff is tilted ($1/a_1\neq 0$) so that magnification can move galaxies across the cut. (In a case where magnification moves galaxies in a direction parallel to the selection boundary, the $\partial P_A/\partial\Delta$ term would be zero and this aspect of our model would not apply.) Figure \ref{fig:selection} illustrates our fiducial selection cut. Under this model, the probability of selection is given by 
\begin{equation}
     \frac{1}{2\pi\sqrt{\det(\cfp)}} \int_{-\infty}^{\infty}\int_{a_0+a_1\mu}^\infty\exp\left(-\frac{1}{2}(\pstr - \overline\pstr)^T\cfp^{-1}(\pstr - \overline\pstr)\right)d(\log_{10}R) d\mu.
\end{equation}
In our fiducial model the number of galaxies that pass selection and fall within the redshift bins we define is about 39,000 (out of 100,000 initially generated).
\\
\section{Emulation Code, Mock Measurement, \& Testing} \label{sec:testing}

We have written a Python code that generates a source galaxy population based on the physical model described above, creates mock catalogs, carries out the estimation procedure, and tests the results for accuracy. Two basic tasks form the building blocks for these procedures. First, the code creates \code{Universe} objects, which contain information about a source population and a foreground lens, as well as a data vector that summarizes the numbers densities and fundamental plane residuals in bins of redshift and angular position. That is, a \code{Universe} contains all the information necessary to probe the mass profile of a lens using the fundamental plane. Second, this code implements the estimator described in section \ref{sec:estimator}. Finally, the code combines these two functionalities to assess the accuracy of the estimator for the lens mass and redshift error parameters by applying the estimation procedure many times with input parameter values close to fiducial parameters. 

\subsection{Creating an observable Universe}
\begin{figure}[h]
\centering
\includegraphics[scale=0.275]{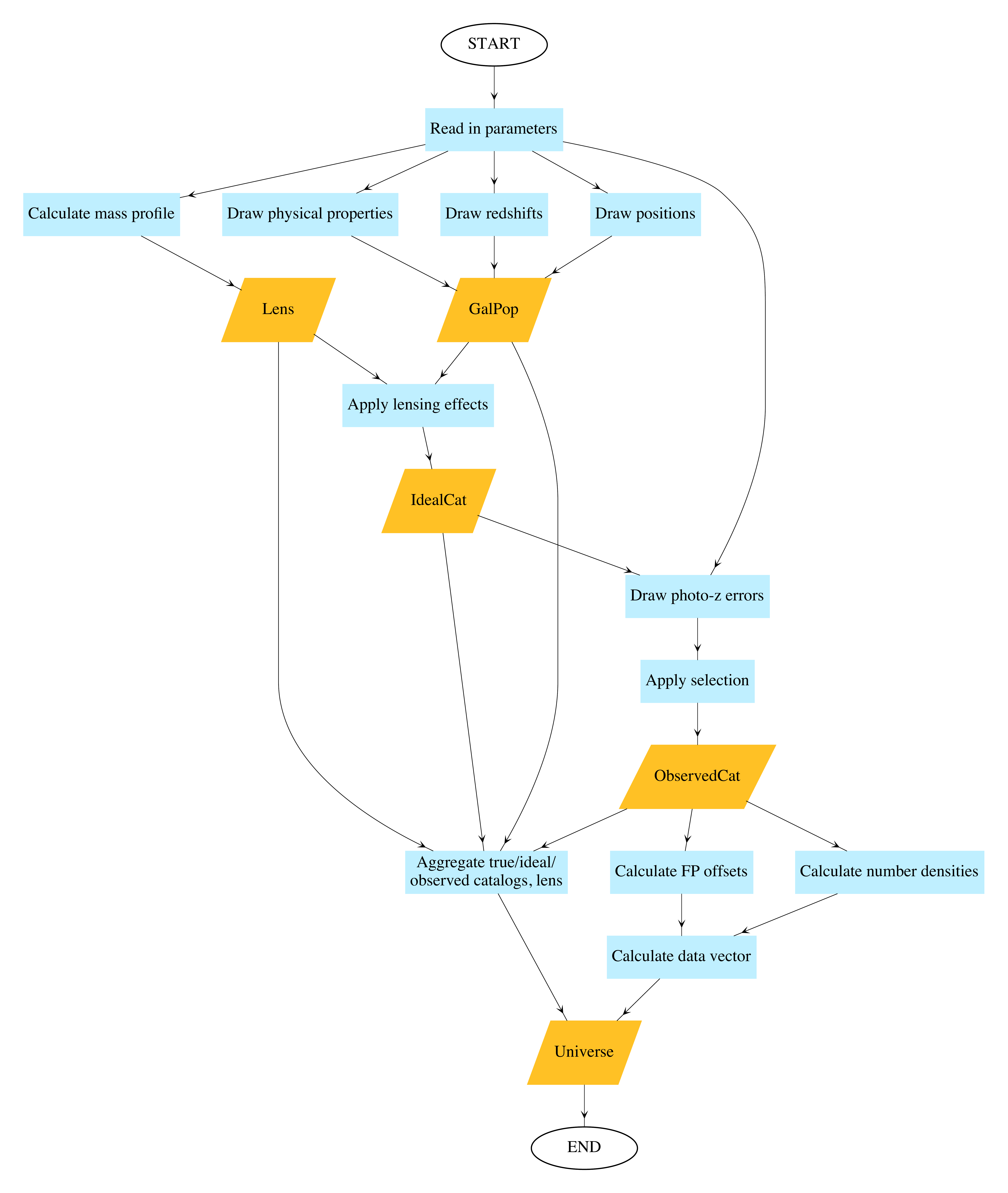}
\caption{Diagram of the catalog emulator code. Rectangles represent procedural steps and parallelograms represent key data products.}
\label{fig:universe_flowchart}
\end{figure}

A \code{Universe} object aggregates several types of catalogs constructed from a single source population, as well as a foreground lens. Furthermore, it provides methods for calculating the data vector of observations from this galaxy population. A schematic for this portion of the code is shown in figure \ref{fig:universe_flowchart}.

\subsubsection{Lens}

A \code{Lens} object is constructed from a virial mass and redshift specified by the user. The halo concentration is determined using the power-law fit from \citet{Mandelbaum:2008aa}\footnote{This halo mass-concentration relation is calibrated to a halo at redshift 0.22, which is above our fiducial value; however, for the purposes of a formalism paper, the relation constitutes a reasonable modeling choice and can easily be altered in future analyses.}. The \code{Lens} object also contains methods for calculating the surface mass density at angular separation $\theta$, as well as $\kappa(\theta,z)$ and other lensing quantities for a given background galaxy, based on the formalism described in section \ref{ssec:lensing}. 

\subsubsection{Catalogs}

We create a \code{GalPop} object that sets true values for the redshifts and angular positions of each galaxy in a population of a specified size. These quantities are drawn from the probability distributions specified in section \ref{sec:physicalmodel}. At this stage, the true angular sizes are calculated and galaxies are sorted into their redshift bins. The size and surface brightness are drawn from a 2-dimensional normal distribution with mean and covariance specified by the user.

An \code{IdealCat} object is constructed from a \code{GalPop} object and a \code{Lens} object. The lensing equations are applied to each galaxy in the \code{GalPop} catalog, which alters their sizes and angular separations. Both the true and lensed quantities are stored in the \code{IdealCat} catalog. 

Once this is done, methods of the \code{ObservedCat} object draw photo-z errors for each galaxy, as described in section \ref{sec:physicalmodel}. Based on these values, systematic effects are applied to the size and surface brightness from the \code{IdealCat} catalog, and the results are stored in the \code{ObservedCat} catalog.

\subsubsection{Universe}

A single \code{Universe} object acts as a container for one set of \code{Lens}, \code{GalPop}, \code{IdealCat}, and \code{ObservedCat} objects. Using the FP covariance and mean provided by the user, we calculate the fundamental plane residual for each galaxy. (As noted in section \ref{ssec:truncated_distributions}, fitting the plane directly from the observed catalog is inaccurate where selection is significant, so the user is assumed to have access to the ``true'' FP fit from another source such as deep-field observations. Tools such as Balrog \citep{Suchyta:2016aa} may also be useful for this purpose.) The average FP residual in each user-specified bin of angular separation and redshift is appended to the average number density in each bin, and this constitutes the final data vector of observations.

\subsection{The estimation procedure}

The code implements the estimation procedure by applying the maximum likelihood estimator described in section \ref{sec:estimator} to the emulated observations contained in a \code{Universe} object. Here, the model $M(\Theta)$ is understood as the output of \code{Universe.data\_vector} when a \code{Universe} object is instantiated from a configuration file containing a given set of input parameters $\Theta$, where the parameters of interest are designated by the user. For example, in this analysis, the target parameters we wish to estimate jointly are the mean $\mu_{z,i}$ for each bin, $\sigma_z$, and $M_H$.

The estimator requires three inputs: a full set of parameters for realizing the \code{Universe} which will provide the mock observations; a set of fiducial parameters; and the covariance of the data vector. In order to obtain the latter, we generate 10,000 \code{Universe} objects from the fiducial parameters and measure the covariance of the resultant data vectors. The other inputs are provided by the user in the configuration file. Although not required by the code, we abide by the general rule that non-targeted input parameters should match the fiducial parameters. From these inputs, we first calculate the linear response of the data vector, corresponding to equation \ref{eqn:linear_response}. This is done by applying small perturbations (of order $10^{-3}$) to the target parameters around fiducial values. We then difference the data vectors of the realized and fiducial \code{Universe} objects to obtain $A$ and apply equation \ref{eqn:estimator}, which yields estimates for the target parameters based on observations from our realized \code{Universe} object. This result constitutes one iteration of the estimation procedure. 

The procedure described above is the most basic implementation of the estimator; in practice, we apply several options in the code for improving the accuracy of the final estimate. First, the linear response of the data vector contains a certain amount of noise if the fiducial and realized \code{Universe} objects being compared are randomly generated - that is, the galaxies in the fiducial \code{Universe} do not have exact counterparts in the realized \code{Universe} whose properties differ by an amount obtainable from the input parameters. Instead, we expect the averaged properties to differ in a predictable way, with some noise stemming from the fact that each population is randomly generated. However, this noise may be suppressed if we require the realized and fiducial \code{Universe} objects to be generated from the same random seed (using the \code{RandomState} feature of the \code{numpy.random} module). For each estimation, we carry out this seed-paired calculation of the linear response a number of times (given by \code{nmlr}, as in table \ref{tab:settings}) and take the mean. This averaging controls noise across multiple instantiations of the same-seed \code{Universe} pairs. Using the mean linear response and the covariance pre-calculated at fiducial values, the estimate is determined from a fully random (i.e. not seed-paired) fiducial-realized set of \code{Universe} objects..

As originally constructed, our estimator is based on a linear expansion around a fiducial model. We therefore use iteration to achieve accurate parameter estimation even for points far from the fiducial model. After an initial estimate, we set the fiducial parameters to this result and redo the procedure 
\newpage 
\clearpage
\begin{tikzpicture}[remember picture,overlay,every node/.style={anchor=center}]
  \node[anchor=north west,xshift=-1cm,yshift=2cm] at (current page.north west)
     {\includegraphics{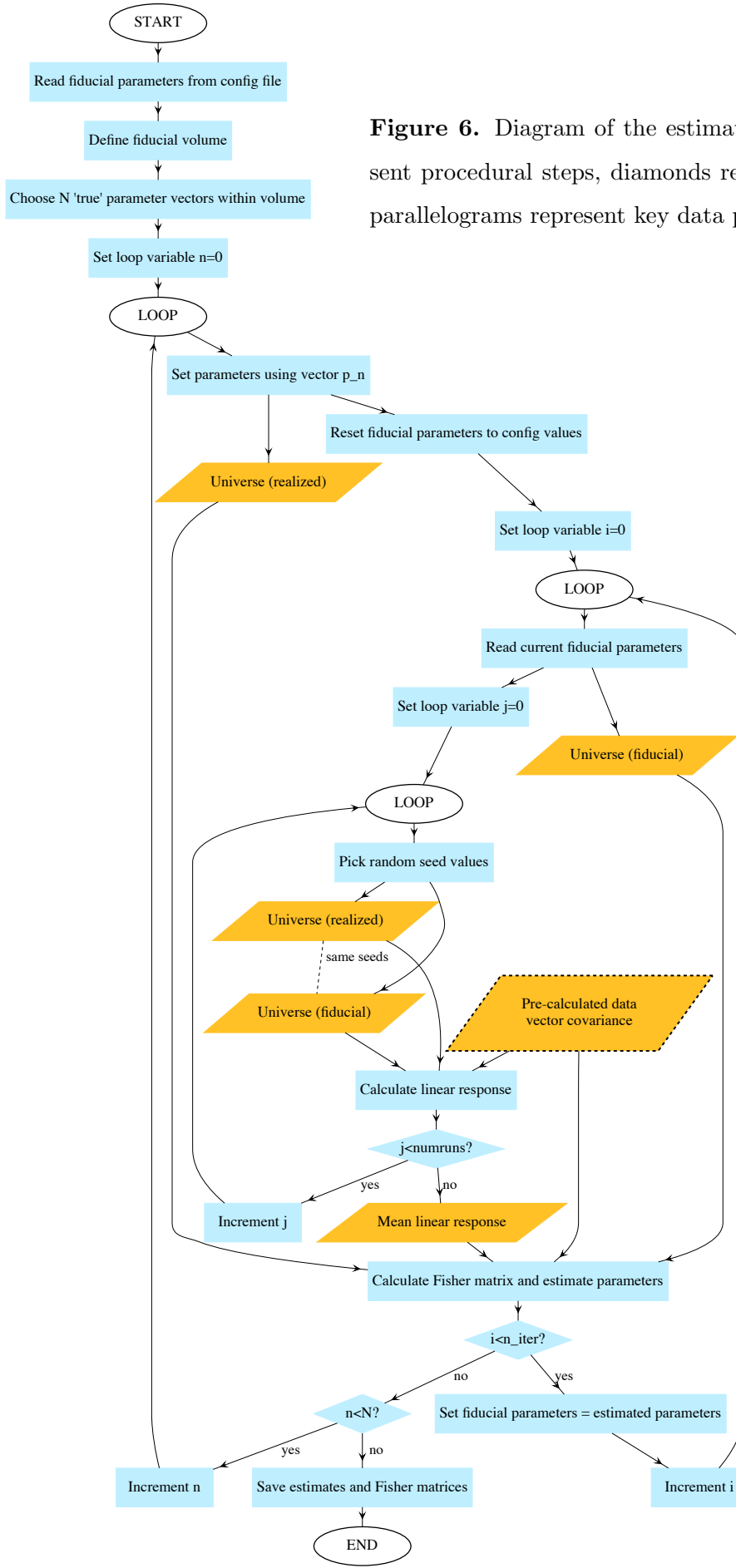}};
  \node [fill = white!30, anchor = north west, minimum height = 4mm, minimum width = 0.20\paperheight, align = flush left,  inner xsep = 1mm, text width=0.40\paperheight-5mm, rotate = 0] at (page cs:-0.1,0.95) {\captionof{figure}{Diagram of the estimation code. Rectangles represent procedural steps, diamonds represent decision points, and parallelograms represent key data products.}\label{fig:estimation_flowchart}};
\end{tikzpicture}
\clearpage
\newpage
\noindent with a newly calculated mean linear response. One iteration beyond the initial estimation is sufficient for our purposes, although the code allows for higher number of iterations (\code{niter} in table \ref{tab:settings}).

\subsection{Verification}
\begin{figure}[h]
    \centering
    \includegraphics[width=\textwidth]{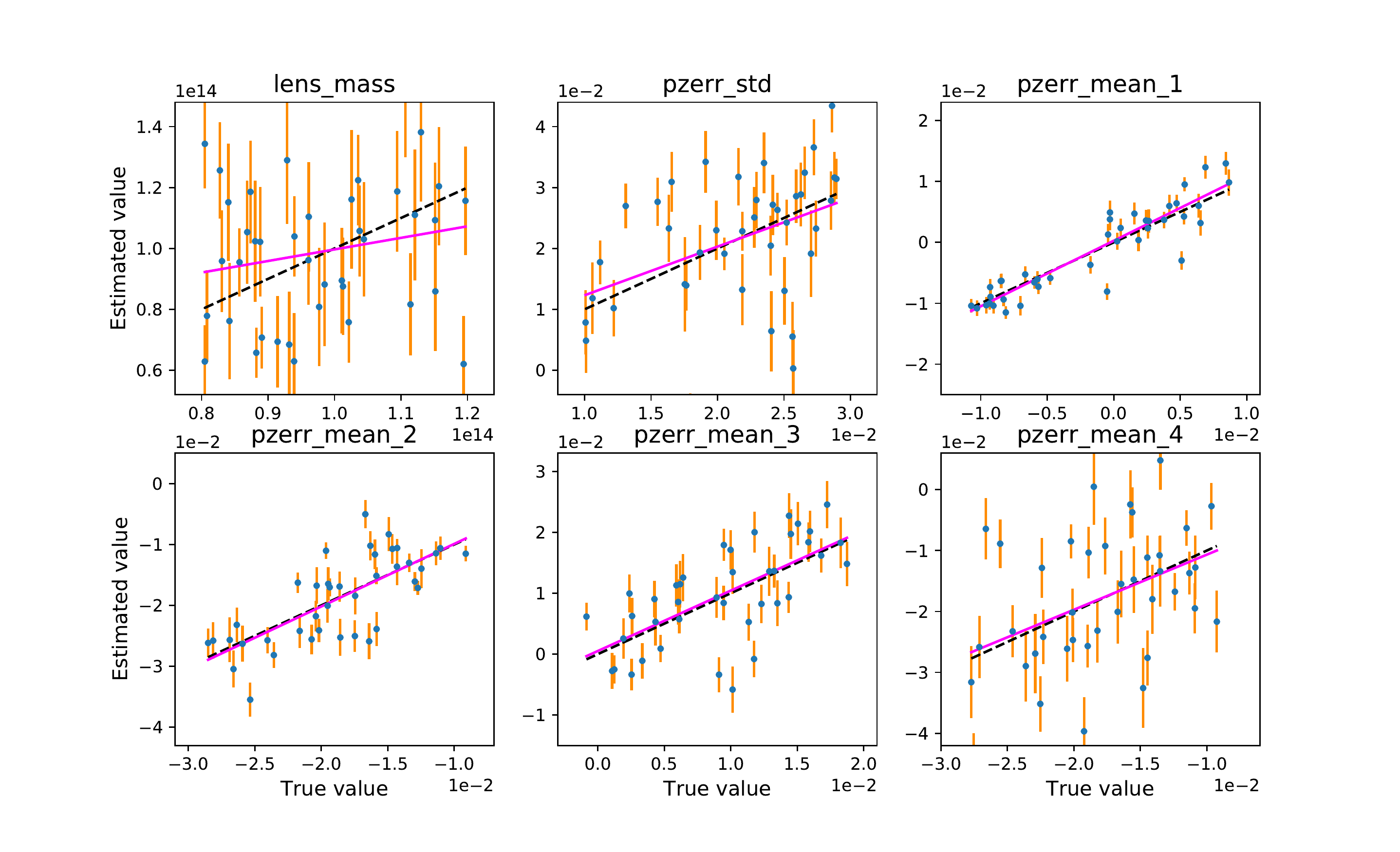}
    \caption{The performance of the estimator for input values near the fiducial parameters for an analysis including FP residuals, number counts, and selection (our standard analysis). Black dashed lines indicates exact correspondence between the input parameter value and the estimated value. Blue points indicate the value returned by the estimator, and orange lines indicate error bars calculated from the corresponding element of the inverse Fisher matrix. Dashed pink lines are linear fits to the estimated values.}
    \label{fig:convergence_tests_selection}
\end{figure}

\begin{figure}[h]
    \centering
    \includegraphics[width=\textwidth]{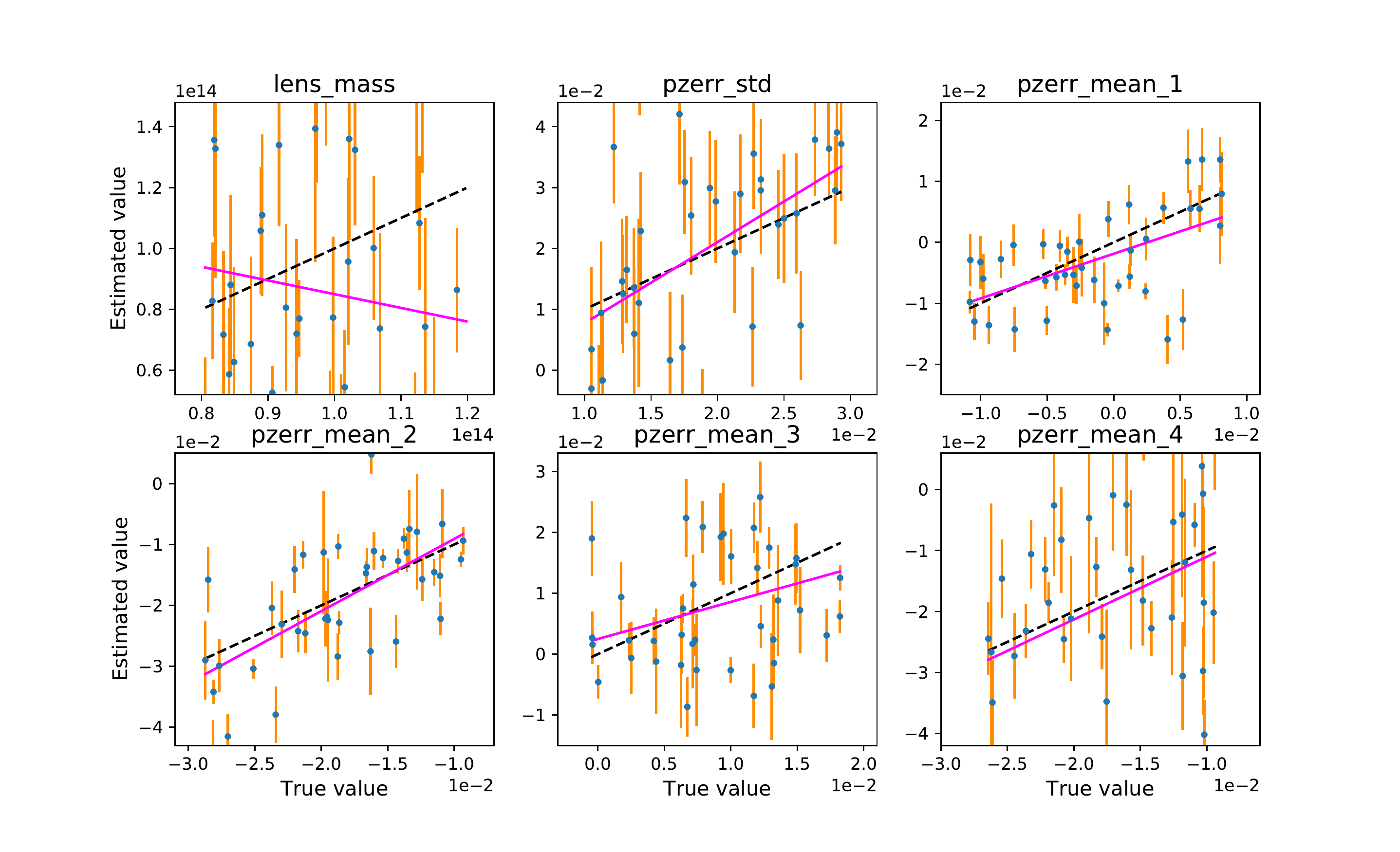}
    \caption{The performance of the estimator for input values near the fiducial parameters in the case where a number-counts-only data vector is employed. See caption of figure \ref{fig:convergence_tests_selection} for description of plot features.}
    \label{fig:convergence_tests_selection_nconly}
\end{figure}

\begin{figure}[!htb]
    \centering
    \includegraphics[width=\textwidth]{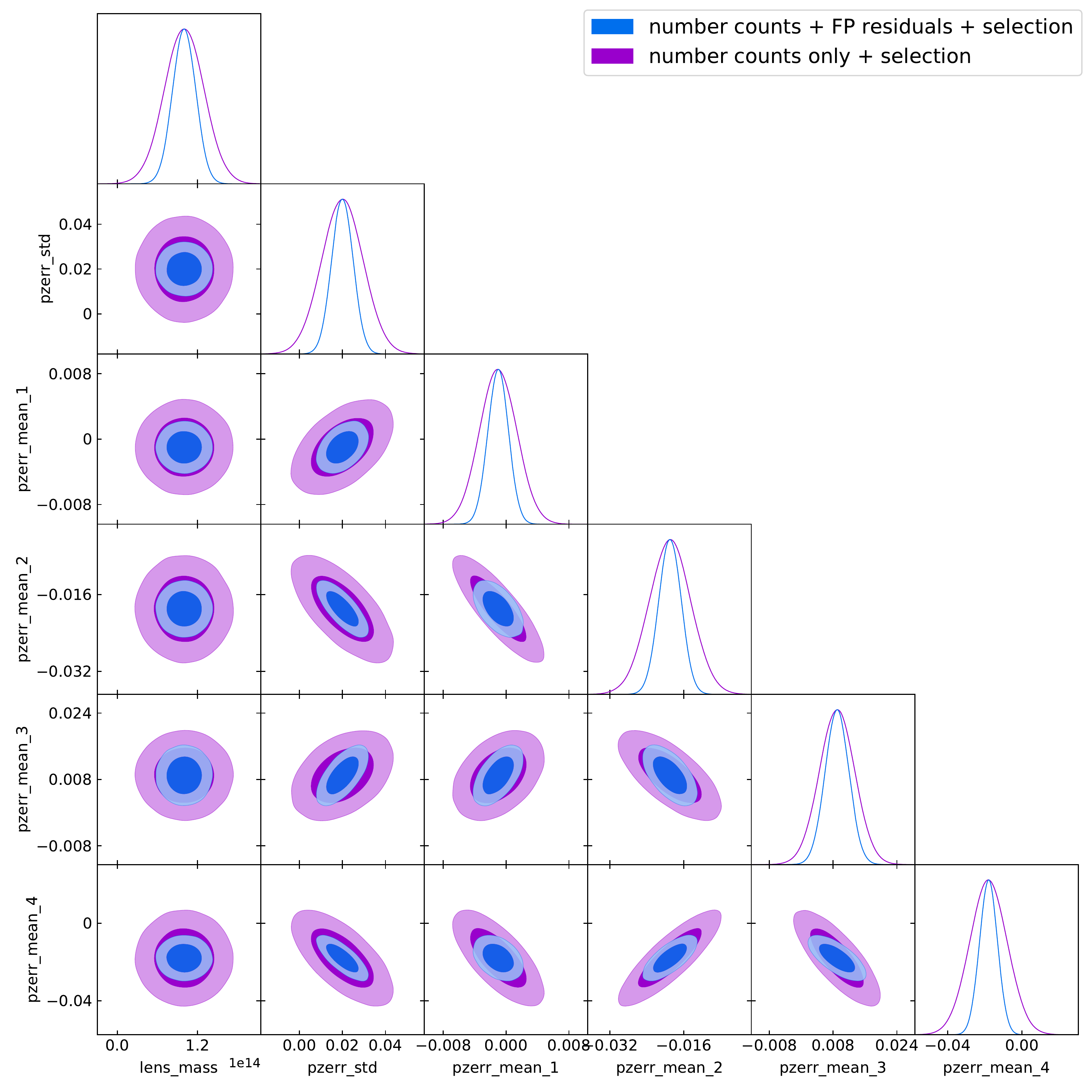}
    \caption{Constraints obtained from the inverse Fisher matrix averaged over $N$ estimations for our standard analysis (blue) and a number-counts-only analysis (purple). Contours are centered on the fiducial values.}
    \label{fig:fisher_combined_nc_vs_fpnc}
\end{figure}

\begin{figure}[!htb]
    \centering
    \includegraphics[width=\textwidth]{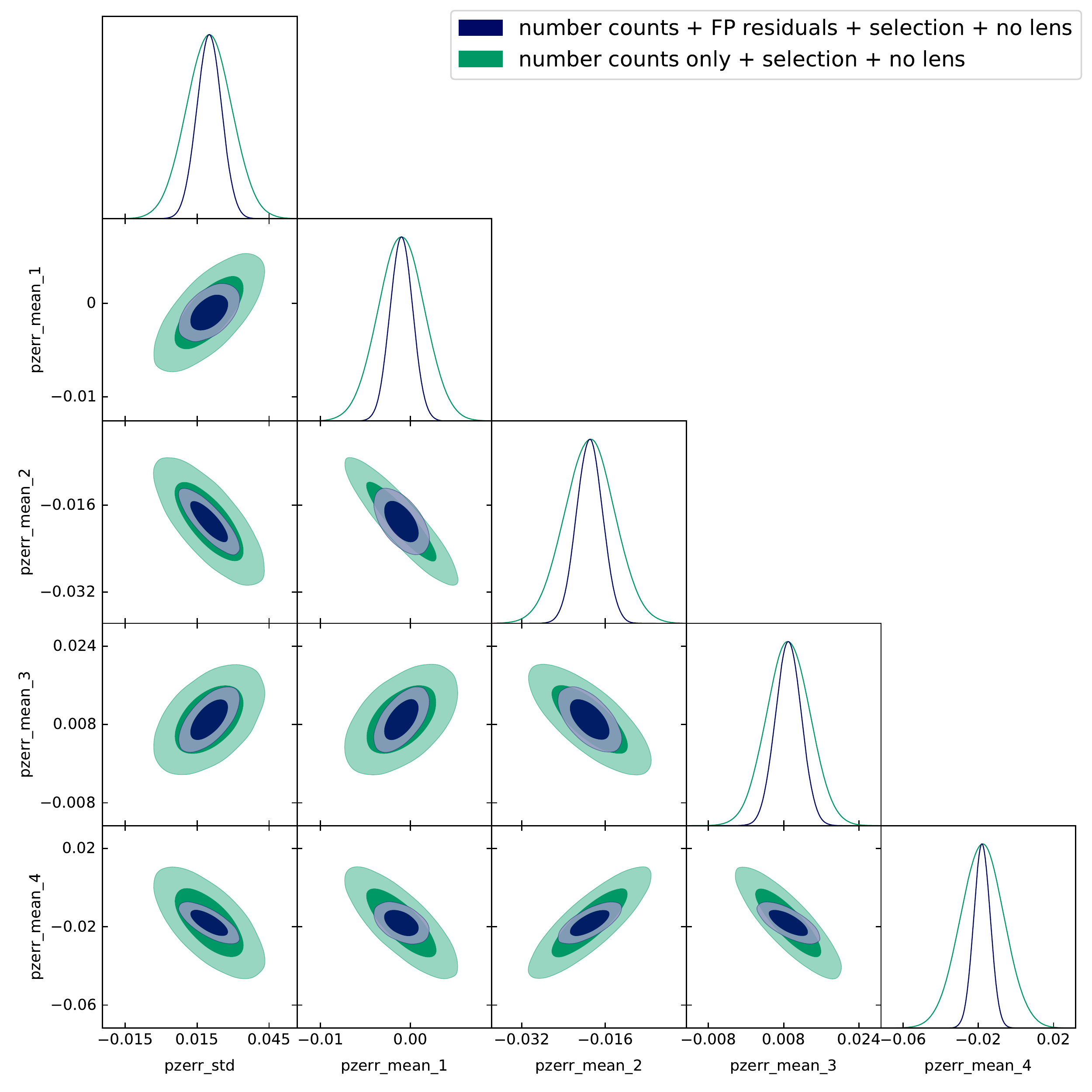}
    \caption{Constraints obtained from the inverse Fisher matrix averaged over $N$ estimations with no lensing included, for the case using a standard data vector (navy) and a number-counts-only data vector (green). Contours are centered on the fiducial values.}
    \label{fig:fisher_combined_nolens}
\end{figure}

\begin{figure}[!htb]
    \centering
    \includegraphics[width=\textwidth]{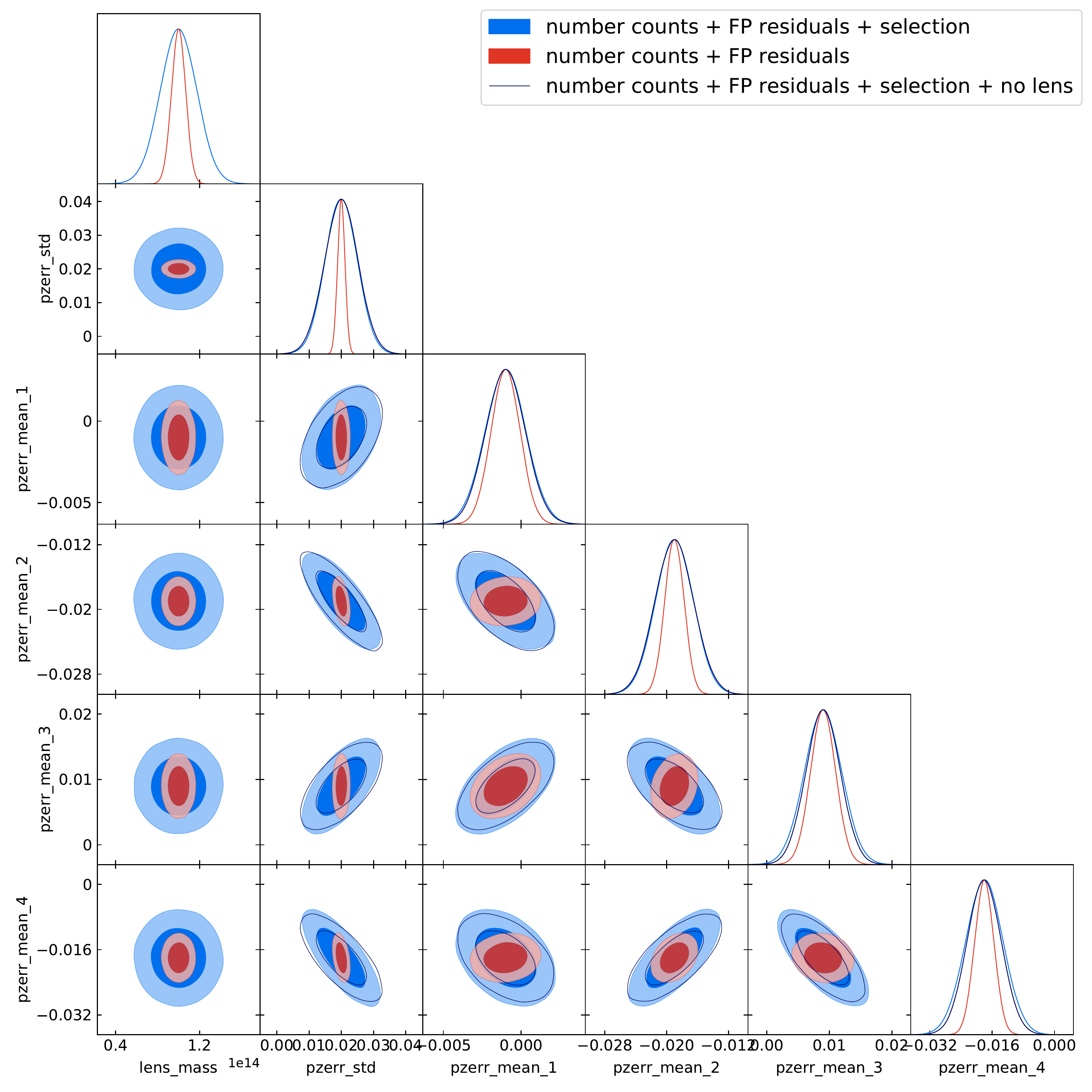}
    \caption{Constraints obtained from the inverse Fisher matrix averaged over $N$ estimations for our standard analysis (blue) and a case without selection effects (red). Unfilled contours correspond to a case with a standard data vector and no lensing. Contours are centered on the fiducial values.}
    \label{fig:fisher_combined_allfp.pdf}
\end{figure}

We verify the accuracy of our procedure for estimating on a single set of observations by tracking its performance on \code{Universe} objects realized from many different parameter values. A schematic of the code for this verification is shown in figure \ref{fig:estimation_flowchart}. The code defines an $L$-dimensional volume, where $L$ is the number of target parameters, around the fiducial values. We then randomly choose a set of $N$ points within the fiducial volume. In our analysis, $L=6$ and $N=40$. The boundaries in each parameter dimension are given by \code{cube\_size} (see table \ref{tab:settings}). For each point, a \code{Universe} object is realized with input parameters set accordingly. We then apply the estimation procedure described above and compare the results to the input values. We run this test for several scenarios, including our standard analysis, a case in which information from the FP residuals is omitted from the data vector and constraints are obtained from number counts only, and a version of our standard analysis that omits lensing altogether and estimates only the photo-z parameters. For each estimation, we calculate an error bar from the inverse Fisher matrix. Figures \ref{fig:convergence_tests_selection}-\ref{fig:convergence_tests_selection_nconly} show these results for the standard and number-counts-only analyses. From the average of the inverse Fisher matrix over $N$ estimations, we plot the contours shown in figures \ref{fig:fisher_combined_nc_vs_fpnc}-\ref{fig:fisher_combined_allfp.pdf}.

\begin{table}[h]
    \centering
    \begin{tabular}{|l|M{60mm}|M{60mm}|}
    \hline
Key & Description & Value \\
\hline
\hline
\code{dres} & $\delta \Delta$ value for calculating selection term in FP residual & 0.01 \\
\hline
\code{data\_vector\_cov\_numruns}  &  Number of instantiations of fiducial universe for calculating covariance                     & 10000 \\
\hline
\code{perturbation\_factor}        &  $\delta\Theta$ value for calculating linear response & 0.003 \\
\hline
\code{target\_params}              &  parameters to estimate & \code{lens\_mass}, \code{pzerr\_std}, \code{pzerr\_mean\_1}, \code{pzerr\_mean\_2}, \code{pzerr\_mean\_3}, \code{pzerr\_mean\_4} \\
\hline
\code{theta\_bins}                &  bins of angular position  & 10 equal-size linear bins over interval (0.01 arcsec, 300 arcsec) \\
\hline
\code{z\_bins}                    &  bins of redshift (chosen to match \citet{Hoyle:2018aa}) & (0.2 , 0.43, 0.63, 0.9 , 1.3) \\
\hline
\code{N}                          & number of estimations & 40 \\
\hline
\code{nmlr}                       & number of linear response calculations per estimation & 20 \\
\hline
\code{niter}                      & number of iterations over fiducial values & 2 \\
\hline
\code{cube\_size} & bounds of the fiducial volume & \code{lens\_mass}: $(\Theta_i-0.2\Theta_i,\Theta_i+0.2\Theta_i)$ \hspace{40pt} other: $(\Theta_i-0.01,\Theta_i+0.01)$ \\
\hline
    \end{tabular}
    \caption{Settings used in this analysis.}
    \label{tab:settings}
\end{table}

%%%%%%%% CONVERGENCE TESTS %%%%%%%%%%

%%%%%%%%%%% FISHER PLOTS %%%%%%%%%%%

\section{Results \& Discussion}\label{sec:results}

In this section, we present the results of the estimation procedure applied to our fiducial model, as implemented in our code. The reader should bear in mind that our modeling choices greatly simplify the messy physical and observational realities of a DES-like survey. Therefore, we emphasize on the relationships among different parts of our model and how they contribute to the final constraints we obtain, rather than the numbers attached to these constraints (presented in table \ref{tab:constraints}). We encourage the reader to view these numbers as the first step towards a meaningful magnification measurement, rather than conclusive results.

From figure \ref{fig:convergence_tests_selection}, we can see that our estimation procedure accurately recovers input values over a wide range around our fiducial values. The performance of the estimator in recovering the parameters describing the mean of the photo-z bins is particularly good, while more noise is present in the estimations of lens mass and the amplitude of the photo-z error.

A version of our analysis in which no FP residuals are used in the data vector (shown in figure \ref{fig:convergence_tests_selection_nconly}) proves significantly noisier and less constraining than our standard analysis. This is evident not only from a comparison of the estimator's convergence to true values in each case, but also from a comparison of the blue and purple contours in figure \ref{fig:fisher_combined_nc_vs_fpnc}, which show that an analysis including FP residuals yields mass constraints about 50\% tighter than an analysis with number counts only. For the photo-z error parameters, the difference is even starker, with the contour sizes for a number-counts-only data vector approximately twice as large. These differences demonstrate that the FP residuals provide a significant boost in the constraining power of this measurement method. (We note that in the standard method for measuring magnification from number counts, the galaxy sample is significantly larger, since it is not limited to elliptical galaxies and can work to lower S/N ratio. The point here is that the FP is adding information that is not in the number counts of the same objects.)

We make a similar comparison in figure \ref{fig:fisher_combined_nolens}, which shows the difference between a number-counts-only and a number-counts + FP-residuals analysis in the case where no lensing is present. In this case, our method probes the photo-z parameters only. The green and navy contours here are nearly identical to the purple and blue photo-z error contours in figure \ref{fig:fisher_combined_nc_vs_fpnc}. This shows that the information constraining the photo-z error parameters does not come from the interaction of magnification with the apparent number density or the FP; rather, the FP residuals and number counts themselves are constraining the photo-z error, given the underlying redshift distribution. Of course in a realistic analysis, we will have to marginalize over the intrinsic redshift distribution and FP parameters, or constrain them from external information. Therefore, we advise the reader {\em not} to read too much into the specific photo-z constraints for these cases; the important points are (i) that the lens mass is very \\
\\
\\
\\
\\
\\
\\
weakly degenerate with these parameters, and (ii) that our framework enables tests such as this to trace the ultimate sources of information.

Figure \ref{fig:fisher_combined_allfp.pdf} compares all cases that incude both number counts and FP residuals: the case including selection effects (the standard analysis), the case without selection, and the case with selection but no lensing. As noted above, the constraints from the lensing-free case, shown by the unfilled contours, adhere closely to those from the standard analysis for the photo-z error parameters. In contrast, the constraints on these parameters are moderately affected by selection, particularly for lens mass and the amplitude of the redshift error. In the absence of selection effects, these parameters are much better constrained than when our fiducial selection boundary is applied. This indicates that the increase in noise from the decreased number of galaxies observed in an incomplete sample dominates over the information gained from magnification moving individual galaxies across the selection boundary (see equations \ref{eqn:likelihood_selection}-\ref{eqn:dpdDelta_term}).

\begin{table}[h]
    \centering
    \begin{tabular}{|l|c|c|c|c|c|}
    \hline
{} & Fiducial value & var & Constraint \\
\hline
\code{lens\_mass}     &       $1.00\times10^{14} M_\odot$   & $2.99\times10^{26}M_\odot^2$ &   17.0\%         \\
\code{pzerr\_std}     &       0.02                          & $2.48\times10^{-5}$ &   24.3\%                  \\
\code{pzerr\_mean\_1} &      -0.001                         & $2.34\times10^{-6}$ &   $\pm 1.50\times10^{-3}$ \\
\code{pzerr\_mean\_2} &      -0.019                         & $6.09\times10^{-6}$ &   $\pm 2.39\times10^{-3}$ \\
\code{pzerr\_mean\_3} &       0.009                         & $1.07\times10^{-5}$ &   $\pm 3.23\times10^{-3}$ \\
\code{pzerr\_mean\_4} &      -0.018                         & $2.35\times10^{-5}$ &   $\pm 4.74\times10^{-3}$ \\
\hline
    \end{tabular}
    \caption{Constraints for each of the target parameters from the standard analysis, averaged over 40 estimations. Column 2 is the mean variance of the target parameter, as obtained from the inverse of the Fisher matrix (see equation \ref{eqn:var_sigma}). In the final column, values prefixed with $\pm$ refer to the mean square root of $\mathrm{var}_j$, and values given as percentages refer to this quantity divided by the fiducial value.}
    \label{tab:constraints}
\end{table}

There are a number of ways to quantify the constraining power of this estimator as implemented here. For example, we can observe that the the mass of the foreground lens for the stacked lens scenario that we're emulating here is constrained to 17\% by our analysis (see table \ref{tab:constraints}). This constraint is derived from the mean variance from the Fisher matrix for each estimation, i.e.
\begin{equation}\label{eqn:var_sigma}
    %\sigma_j = \frac{1}{N}\sum_{i=1}^N \left({\rm F}^{-1}_i\right)_j^{1/2} {\rm~~~and~~~}
    \mathrm{var}_j = \frac{1}{N}\sum_{i=1}^N \left({\rm F}^{-1}_i\right)_j, 
\end{equation}
where where F is the Fisher matrix and $j$ runs over the target parameters. However, it is also instructive to generate a quantity similar to $\sigma^2_\gamma$, the shape noise in shear analyses. The quantity in our analysis most directly comparable to the variance of the ellipticity in a weak lensing sample is the variance $\sigma^2_{\kappa, \rm FP}$ of the fundamental plane residuals, as calculated directly from the mock observations. For our fiducial universe we obtain $\sigma_{\kappa,\rm FP} =0.358$. 

While $\sigma^2_{\kappa, \rm FP}$ is the measured quantity in our catalog most directly analogous to shape noise, the constraining power of our approach is perhaps better expressed by the effective measure $\sigma_{\kappa,\mathrm{eff}}^2$, which ties our constraint on halo mass to a constraint on $\kappa$ itself. We make the definition 
\begin{equation}
    \sigma_{\kappa,\mathrm{eff}}^2 \equiv \sigma_M^2\sum_u\frac{\partial\kappa}{\partial M_H}\bigg\vert_u^2\,,
\end{equation}
where $u$ indexes the source galaxies and $\partial\kappa/\partial M_H$ may be calculated from the lensing formalism in section \ref{ssec:lensing}. For our standard analysis, we obtain $\sigma_{\kappa,\mathrm{eff}}=0.250$. In comparison, the DES Y1 weak lensing catalog has a shape noise $\sigma_\gamma\approx0.28$ \citep{Zuntz:2018aa}\footnote{Note that the number we quote here as $\sigma_\gamma$ is $\sigma_e$ in the notation of \citet{Zuntz:2018aa}, since they use $\sigma_\gamma$ to refer to a density-weighted quantity (see their table 5).}. %We would expect $\sigma_\kappa$ to exceed $\sigma_\gamma$ in part because the distribution of galaxy sizes has more intrinsic scatter than the distribution of galaxy shapes. It is worth noting that while $\kappa$ has more noise, it also has a steeper profile than $\gamma$ on small scales (see figure \ref{fig:kappagamma}), so generally speaking it is not a foregone conclusion that a noisy magnification measurement is less useful than a shear measurement.

These calculations provide a useful reference point for situating our analysis within the wider weak lensing landscape. However, we again emphasize that the numbers attached to these constraints are specific to the modeling choices we have made. In many cases, we have chosen idealized physical implementations, so that we can trace the flow of information through the estimation process. In particular, we have treated the fundamental plane as a standard ruler, which is quite a simplification of reality. However, the virtue of this approach is that it allows us to rigorously characterize the effect of each model component on our final constraints. Our ability to simultaneously constrain the photo-z error parameters and the foreground mass, and to articulate where in the estimation process these constraints arise, is a particular advantage. 

\section{Conclusions \& Future Work}\label{sec:conclusions}

The framework laid out in this paper provides the foundation for a galaxy-magnification cross-correlation measurement in a photometric galaxy survey. One goal of such a measurement is to integrate magnification information into the ``$N\times 2$-point'' joint analysis that constitutes the standard pipeline for cosmological parameter estimation. A measurement of sufficient precision and accuracy for such an analysis will require careful modeling of physical and observational effects beyond the assumptions we make in this paper. Our method is meant to be extended and refined, and we plan to add a number of features that will lead to a measurement on DES data in the near future.

Improving our fundamental plane model is the first priority. One possible improvement would be to modeling the plane in 3D by adding a concentration parameter is one option. However, as discussed previously, we anticipate that concentration will be difficult to parameterize in a DES-like survey, and may require calibration by a spectroscopic sample. Therefore our priority is to explicitly model the FP over a range of redshifts and galaxy environments. This will allow us to compare each galaxy to its ``local'' FP, rather than a global FP measured from the entire sample, reducing scatter and bias due to correlations in the FP residuals. As a further extension, we plan to model the FP in colorspace, in order to take full advantage of the multi-band information available in a photometric survey. Incorporating colors means that our model will be sensitive to dust, so we will include a dust model for our foreground as well.

We will also need to add complexity to our model on the observational side. For example, we have chosen a simple cutoff in size and surface brightness as our selection function, but we expect the selection function in real data to prove less tractable. We plan to make heavy use of Balrog \citep{Suchyta:2016aa}, a tool that injects fake galaxies into DES imaging and runs them through the image processing pipeline. This will allow us to create a survey-specific model for selection, as well as a number of other systematics.

Beyond these immediate next steps, we anticipate that the FP framework for measuring magnification will serve as a useful basis not only for cosmological parameter estimation, but also for a variety of applications to galaxies and their environments.

\section{Acknowledgements}
During the preparation of this work, JF and CMH were supported by the Simons Foundation award 60052667; US Department of Energy award DE-SC0019083; and NASA award 15-WFIRST15-0008. Parts of this research were carried out at the Jet Propulsion Laboratory, California Institute of Technology, under a contract with the National Aeronautics and Space Administration.

\software{
        Astropy \citep{Astropy-Collaboration:2018aa};  
        GetDist \citep{getdist};  
        IPython \citep{ipython};  
        Jupyter \citep{jupyter};  
        Matplotlib \citep{matplotlib};  
        Numpy \citep{numpy};  
        pandas \citep{pandas};  
        PyGraphviz \citep{pygraphviz};  
        SciPy \citep{scipy};  
        tqdm \citep{tqdm}
        }

\clearpage
\bibliographystyle{aasjournal} 
\bibliography{references}

\end{document}